\documentclass[preprint,10pt]{elsarticle}

 \usepackage{graphicx}

\usepackage{amsmath}
\usepackage{amssymb}

\journal{submitted}
\begin{document}

\title{Linear regression for numeric symbolic variables: a least
squares approach based on Wasserstein Distance}

\author{Antonio~Irpino}

\ead{antonio.irpino@unina2.it}

\author{Rosanna~Verde}

\ead{rosanna.verde@unina2.it}

\address{Dipartimento di Studi Europei e Mediterranei, Seconda Universit\'{a}
degli Studi di Napoli, Viale Ellittico 31, 81100, Caserta, Italy}
\begin{abstract}
In this paper we present a linear regression technique for numerical modal symbolic
data. The observed variables can be histogram-valued data, empirical distributions or empirical estimates of parametric distributions that are considered modal data according to the Symbolic Data Analysis approach.
In order to measure the error
between the observed and the predicted distributions, the $\ell_2$
Wasserstein distance (known also as Mallow's distance) is proposed. Some properties of such metric are
exploited to predict the response variable as direct linear combination
of other independent modal variables. Using a particular decomposition of the distance, we propose a Least Square method for the estimates of the model parameters. Measures of goodness of fit are also proposed and discussed. We propose the technique in a description functions because the inferential properties of the symbolic variables are, up to now, not yet well studied and is still a new field of research.
An application on real data shows the main advantages of using the proposed method against the existing ones.
\end{abstract}
\maketitle

\section{Introduction}

In this paper we present a linear regression model for modal symbolic
data in the framework of Symbolic Data Analysis (SDA).

Since the first pioneering papers of Edwin Diday, SDA was become a
new statistical field of research gathering contributions from different scientific communities:
statistics, knowledge extraction, machine learning, data mining.
Three European projects supported the systematic developments of the
main methodological contributions and a wide collection of SDA methods
is available in three reference books (\citet{bock2000analysis,billard2006symbolic} and
\citet{diday2008symbolic}). Moreover, many other papers have been published
on Journals and Conference proceedings and currently SDA is almost
always in the list of the topics of the most relevant international
data analysis conferences. The basic ideas of SDA consists of analyzing
data which are described by set-valued variables. Symbolic data refer
to a description of a class, a category, a set of individuals and,
more generally, they correspond to a description of a ``concept''.
Examples of symbolic data are: football teams (classes of individuals);
species of animals (categories); towns (concepts). Differently from
the classic data where each ``punctual''
observation assumes only one value for every variable, symbolic data
take multiple values, such as intervals of continuous variables, different
categories of nominal variables, empirical distribution or probability
functions.

Symbolic data are receiving more and more attention because they are
able to summarize huge sets of data, nowadays available in large databases
or generated in streaming by smart meters or sensors.

A peculiarity of Symbolic data is that they allow to keep the variability
in the data description, considering the interval of values that each
observation can assume or the distribution of values. In this way,
the methodological approaches developed in this context of analysis
must guarantee this information is preserved.

Symbolic Data Analysis methods generalize multivariate data analysis
to that new kind of data and they can be classified at least in three
categories, according to the \emph{input data} - the \emph{method} - the \emph{output data}, as follows:
\begin{description}
\item [{\emph{symbolic-numerical-numerical}:}] symbolic data in
input are transformed into standard data in order to apply classic
multivariate techniques. The results are classic data. For example,
a dissimilarity between interval data is computed considering only the bounding values (minimum and maximum) of the intervals. That is a standard dissimilarity between punctual data and the measure is a single value.
\item [{\emph{symbolic-numerical-symbolic}:}] symbolic data in
input are analyzed according to a classic multivariate techniques
and the results are symbolic data. For example, intervals are transformed
in mid-points and radii, a classic analysis (e.g. linear regression)
is performed on these data but the results (e.g. predictive variable)
are furnished in terms of intervals by reconstructing them from the
estimated mid-points and radii.
\item [{\emph{symbolic-symbolic-symbolic}:}] symbolic data in input
are transformed using generalization/specialization operators and
the results are symbolic data. For example, symbolic data represented
by intervals, histograms or distributions are aggregated in homogeneous
classes through a clustering methods using a criterion of homogeneity
which takes into account the characteristics of the data (internal
structure). The results are classes of symbolic data expressed by
the same kind of variables of the symbolic input data.
\end{description}
Most part of the SDA techniques are \emph{symbolic-numerical-symbolic}, where the input and output data are symbolic and the
methods are generalization of the classic data analysis methods to
this new kind of data. Considerable contributions have been given in retrieving
variability information of the symbolic data through graphic representations
of the results and tools for the interpretation. A large overview
of the SDA methods is in \citet{bock2000analysis} and in \citet{diday2008symbolic}.
The linear regression models proposed in this context of analysis
have been introduced to study the structure of dependence of a response
\emph{symbolic} variable from a set of independent
or explicative variables of the same nature.

The first proposals were regression models for interval data as extension
of the linear dependence model for numerical variables. \citet{Bi_Di_REG_00}
proposed a non-probabilistic approach based on the minimization
of a criterion like the sum of squared error for the parameters estimation.
The method consists in a regression model on the mid-points of the
intervals and it makes reference to the basic statistics (mean, variance,
correlation) introduced by \citet{BER_GOUP_2000} for interval data.
\citet{Neto_Dec04}, \citet{Neto_DeC08} and \citet{Neto_dec2010} improved
the performance of the linear method adding the ranges of the intervals to the mid-points
information in order to include the variability expressed by the interval
data. Therefore, the authors show the differences between this model and
 two regression models on the bounds of the intervals. \citet{Neto_Dec04} introduced an order constrained on the bounds of the
predicted intervals in the model estimation process in order to guarantee the
coherence between the observed and the predicted response variable.
To overcome the problem of possible inversion of the bounds of the
predicted intervals  \citet{Neto_DeC08} suggested a non
linear regression model on the mid-points and the ranges of the interval.
\citet{billard2006symbolic} presented at first a regression model for
histogram variables. This approach is based on the basic statistics:
mean, variance and correlation defined by \citet{BER_GOUP_2000} for intervals when they
are assumed as random variables uniformly distributed. According to
this approach the fitted regression model and their predicted values
are generally single valued. The authors leave as an open problem output
predicted values as symbolic data.

\citet{VE_IRP_2010_OLS} proposed a simple linear regression model
which allows to estimate a histogram response variable as linear transformation
of another independent histogram variable. The main idea is to propose
a suitable metric to measure the sum of squared errors between the
observed and predicted multi-valued data (histograms or distributions).
The Wasserstein distance \cite{Wasse69} $\ell$ is a distance function defined
between the probability distributions of two random variables $X$
and $Y$, on a given metric space $M$. The minimal $L_{1}$-metric
$\ell_{1}$ had been introduced and investigated already in 1940 by
Kantorovich for compact metric spaces \citet{Kanto40}. In 1914 Gini introduced the
$\ell$ metric in a discrete setting on the real line \cite{Gini14} and Salvemini
(in the discrete case) \cite{Salve43} and Dall'Aglio  (in the general case) \cite{DellAg56}
proved the basic representation of $L_{p}$ norm $\ell_{p}$ between
the quantile functions of the two random variables. \citet{Mallows_72}
introduced the $\ell_{2}$-metric in a statistical context. Moreover,
starting from Mallows' work  \citet{BiFree81} described topological
properties and investigated applications to statistical problems as
the bootstrap. They introduced the notion Mallows metric for $\ell_{2}$.
So the $L_{p}$-metric $\ell_{p}$ was invented historically several
times from different perspectives. Historically the notion of Gini
\textendash{} Dall' Aglio \textendash{} Kantorovich
\textendash{} Wasserstein \textendash{} Mallows metric would be correct
for this class of metrics.

This measure that, we refer to henceforth as Wasserstein metric and that
was already proposed by the authors in Clustering methods for
interval \cite{Irpino20081648} and histogram data \cite{citeulike:9479861,citeulike:7031164},
seems particularly adapt in this context. According to this distance
function, we study the dependence relationship of the histogram response
variable from the explicative one considering the respective quantile
functions.

\citet{dias_brito_reg_2011} referring to this last approach proposed
a linear regression model for histogram data, directly interpreting
the linear relationship between quantile functions. In the multiple
regression model, as we will show in the present paper, one of the
main problem is OLS cannot guarantee all the estimated parameters
are positive. It is worth noting that the predicted response variable
is again a quantile function only if it is a linear combination of
quantile functions with positive coefficients. In order to overcome
such inconvenience, \citet{dias_brito_reg_2011} proposed to introduce the
so called \emph{symmetric quantile distributions} in the model as new predictor
variables. However, the meaning of these new entities is not easy
interpretable. In the same paper a new measure of goodness-of-fit
associated to the proposed model is also introduced.

Differently, our proposal is to exploit the properties of a decomposition
of the Wasserstein distance proposed by \citet{DBLP:conf/f-egc/2007},
that is used to measure the sum of squared errors and rewrite the
model splitting the contribution of the predictors in a part depending
from the averages of the distributions and another depending from
the centered quantile distributions. The parameters associated to
the predictors, constituted by the averages of the distributions,
can be indifferently positive or negative because they effect only
on the shift of the predicted quantile distribution. The authors already
demonstrated for the simple regression model \citet{VE_IRP_2010_OLS}
that this leads to guarantee the positiveness of the parameter associated
to the centered quantile function of the only predictor. However, in
the multiple regression model this solution is not guaranteed, so that it needs to force the non negativity of the
parameters by a constrained (Non Negative) Least Squared algorithm.
The rest of paper is organized as follows: in section \ref{sec:2} symbolic data are presented according to the definition given in \citet{bock2000analysis} and \citet{diday2008symbolic} books;
in section \ref{sec:OLS} regression models for Numerical Probabilistic (Modal) Symbolic Variables are introduced and details on the new proposal are furnished; in section \ref{sec:The-goodness-of} some goodness of fit indices are proposed; in section \ref{sec:apply} applications on real data are presented in order to corroborate the procedure.

\section{Numerical symbolic data}\label{sec:2}

Symbolic data allow to describe concepts, individuals or classes of
individuals, by means of multiple values for each descriptor (variable).
The term\emph{ symbolic variable} was coined in order to introduce
such new set-valued descriptions. In a classic data table ($n\times p$
individuals per variables) each individual is described by a vector
of values, similarly, in a \emph{symbolic data table} each individual
is described by a vector of set-valued descriptions (like intervals
of values, histograms, set of numbers or of categories, sometimes
equipped with weights, probabilities, frequencies, an so on). According
to the taxonomy of symbolic variables presented in \citet{bock2000analysis}
and recalled by \citet{Noirh_Brito_2011}, we may consider as numerical
symbolic variables all those symbolic variables whose support is numeric.

Given a set of $n$ individuals (concepts, classes) $\Omega=\left\{ \omega_{1},\ldots,\omega_{n}\right\} $ a
\emph{symbolic variable} $X$ with domain $D$ is a map
\[
X:\Omega\rightarrow D  \quad  X(\omega_{i})\in D.
\]

The different kinds of variable definitions depend on the nature
of $D$. Considering only numerical domains, we can define the
following symbolic variable:
\begin{description}
\item [{{Classic~Variable}}] It is observed when $D\subseteq\mathbb{R}$, (where $\mathbb{R}$ denotes the set of real numbers) i.e.
each individual $\omega_{i}$ is described by a single numeric value
for the variable $X$;
\item [{{Interval~Variable}}] It is observed when $D\subseteq \mathbb{IR}$,
where $\mathbb{IR}$ is the set of all intervals of real numbers $\left[a,b\right]$
where $a,b\in\mathbb{R}$ and $a\leq b$;
\item [{{Modal~Variables}}] According to \citet{bock2000analysis}
the domains of \emph{Modal Variables} are sets of mappings. Considering
different kinds of mapping, several kinds of \emph{Modal Symbolic
Variables} can be defined. let us consider $D\subseteq M$ where $M_{i}\in M$
is a map $M_{i}:S_{i}\rightarrow W_{i}$, such that for each element
of the support $s_{i}\in S_{i}$ it is associated $w_{i}=M_{i}(s_{i})\in\mathbb{R}^{+}$.\\
 If $M_{i}(s_{i})$ has the same properties of a random variable (i.e.
$\int_{s\in S_{i}}w(s)ds=1$, or $\sum_{s\in S_{i}}w_{s}=1$, ), $X$
can be defined as a \emph{Numerical Probabilistic} (Modal) \emph{Symbolic
Variable} (NPSV) and $M_{i}(s_{i})$ can described through a probability
density function $f_{i}(x)$. Particular cases of such data arise
when the generic individual $\omega_{i}$ is described by a model
of random variables, a histogram, an empirical frequency distribution.
In this paper we refer only to such kinds of data that we call
\emph{Numerical Probabilistic Symbolic Data} (NPSD), that are in the
domain of \emph{Numerical Probabilistic }(Modal) \emph{Symbolic Variables}.
\end{description}
For example, if $\omega_{i}$ for variable $X$ is described by a normal distribion with parameters
$\mu_{i}$ and $\sigma_{i}$, we may describe it by its probability
density function (\emph{pdf}) $f_{i}(x)$ as follows:
\[
X(\omega_{i})=f_{i}(x)=\{N(\mu_{i},\sigma_{i})\}.
\]

Using the same notation, and according to \citet{BER_GOUP_2000}
and \citet{billard2006symbolic}, we may consider interval data as
a particular case of NPSD, where the \emph{pdf} is uniform. Given
an interval description of $\omega_{i}$ as $X(\omega_{i})=[a_{i},b_{i}]$,
we may rewrite the same description in terms of NPSD as:
\[
X(\omega_{i})=f_{i}(x)=\{U(a_{i},b_{i})\}.
\]

Histogram data are a particular case of NPSD, where, given the generic
individual $\omega_{i}$, a set of disjoint $K_{i}$ intervals $I_{ki}=[a_{ki},b_{ki}]\: k=1,\ldots,K_{i}$
and a set of positive $K_{i}$ weights $w_{ki}$ such that $\sum_{k=1}^{K_{i}}w_{ki}=1$
, its description for the NPSV $X$ is :
\[
X(\omega_{i})=f_{i}(x)=\left\{ \left(I_{1i},w_{1i}\right),\ldots,\left(I_{ki},w_{ki}\right),\ldots,\left(I_{K_{i}i},w_{K_{i}i}\right)\right\} .
\]

Also in this case it is possible to define a \emph{pdf} for each histogram
data as proposed by \citet{citeulike:9479861}, considering a histogram
as a mixture of uniform \emph{pdf's.}

In another way an interval can be  treated as a histogram
with $k=1$, such that $I_{1i}=[a_{i},b_{i}]$ and $w_{1i}=1$.

Similarly, classic data (single valued numerical data), can be considered
as NPSD, where the description is a \emph{Dirac delta function}
or as histogram data with one thin ($a_{i}=b_{i})$ interval.

\paragraph*{Notation and definitions}

Let $X_1, \ldots, X_j, \ldots X_p$ and $Y$ be the independent and dependent NPSV's observed on a set $\Omega$
of $n$ individuals (concepts or classes). We denote with:

\begin{itemize}
  \item $f_{i}(x_{j})$ and $f_i(y)$  the empirical or theoretical \emph{probability density functions} ($pdf'$s), i.e. NPSD describing the $i-th$ individual (for $i=1, \ldots,n$);
  \item $F_{i}(x_{j})$ and $F_{i}(y)$ the \emph{cumulative distribution functions} ($cdf$'s);
  \item $x_{ij}(t)=F_{i}^{-1}(x_{j})$ and $y_{i}(t)=F_{i}^{-1}(y)$ the \emph{quantile functions} ($qf$'s);
  \item $\bar{x}_{ij}$, $\bar{y}_{i}$ and $s_{ij}$, $s^y_i$  \footnote{$\bar{x}_{ij}=\int\limits _{0}^{1}x_{ij}(t)dt$, $\bar{y}_{i}=\int\limits _{0}^{1}y_{i}(t)dt$ and $s_{ij}=\sqrt{\int\limits _{0}^{1}x_{ij}^{2}(t)dt-\bar{x}_{ij}^{2}}$, $s_{i}^y=\sqrt{\int\limits _{0}^{1}y_{i}^{2}(t)dt-\bar{y}_{i}^{2}}$}, the \emph{means} and the \emph{standard deviations} of the $x_{ij}(t)'s$ and of $y_i(t)$ respectively. They are real numbers;
  \item $\bar{x}_{j}(t)=\frac{1}{n} \sum_{i=1}^n x_{ij}(t)$, $\bar{y}(t)=\frac{1}{n} \sum_{i=1}^n y_{i}(t)$ the means of the sets of n distributions $x_{ij}(t)$ and $y_i(t)$, i.e. the \emph{baricenter distributions};
 \item $\bar{x}_{j}=\frac{1}{n} \sum_{i=1}^n \int\limits _{0}^{1}{x}_{ij}(t)dt=\frac{1}{n} \int\limits _{0}^{1}\bar{x}_{j}(t)dt$, $\bar{y}=\sum_{i=1}^n \int\limits _{0}^{1}\bar{y}(t)dt=\bar{y}=\int\limits _{0}^{1}\bar{y}(t)dt$ the means of the distribution means of the $x_{ij}(t)$'s and $y_i(t)$, or equivalently the means of the baricenter distributions of the $x_{ij}(t)$'s and $y_i(t)$. They are real numbers;
  \item $x_{ij}^{c}(t)=x_{ij}(t)-\bar{x}_{ij}$ the \emph{centred quantile function} , i.e. the quantile function
shifted by $-\bar{x}_{ij}$.
\end{itemize}

\section{OLS linear regression for NPSD}\label{sec:OLS}

Given $X_{1},\ldots,X_{p}$ $p$ explicative
NPSV's and a dependent NPSV $Y$ observed on a set $\Omega$, the aim is to fit the parameters of a linear regression function $\phi(X)$. Denoted with  $\mathbf{X}$ and $\mathbf{Y}$ respectively the matrix collecting the observed values of the $X_{j}$ explicative NPSV's and  the vector of the observed values of the predictive  NPSV  $Y$, we write the regression model as follows:
\begin{equation}
\mathbf{Y}=\phi(\mathbf{X})+\varepsilon
\end{equation}

As for classic data, and according to the definitions of NPSD's,
the model is fitted starting from the following symbolic data table,
where instead of a matrix of scalar values, we have a matrix of NPSD's:
\begin{equation}
[\mathbf{Y}|\mathbf{X}]=\left[\begin{array}{c|ccccc}
f_{1}(y) & f_{1}(x_{1}) & \cdots & f_{1}(x_{j}) & \cdots & f_{1}(x_{p})\\
\cdots & \cdots & \cdots & \cdots & \cdots & \cdots\\
f_{i}(y) & f_{i}(x_{1}) & \cdots & f_{i}(x_{j}) & \cdots & f_{i}(x_{p})\\
\cdots & \cdots & \cdots & \cdots & \cdots & \cdots\\
f_{n}(y) & f_{n}(x_{1}) & \cdots & f_{n}(x_{j}) & \cdots & f_{n}(x_{p})
\end{array}\right].\label{eq:TAB_SYM}
\end{equation}

Two main approaches for the estimation of the parameters of linear
regression model have been proposed when the symbolic data are histograms.
Starting from the elementary statistics proposed by \citet{BER_GOUP_2000},
in a first approach \citet{billard2006symbolic} introduced an extension
of the classic OLS (Ordinary Least Squares) linear regression model
to the histogram-valued variables. A second group of approaches is
based on the use of the quantile functions (which is biunivocal to
a \emph{pdf}) of the NPSD and of the Wasserstein distance for defining
the sum of square errors in the OLS problem. The idea behind the latest
approaches is to predict a quantile function after having observed
a set of quantile functions as predictors.

\subsection{The Billard-Diday model \label{sub:The-Billard-Diday-model}}

According to \citet{billard2006symbolic}, the regression model which
expresses a linear relationship between a set of predictors and a response
histogram variable is based on the assumptions of \citet{BER_GOUP_2000},
they consider a histogram as the representation of a cluster of individuals.
A second implicit assumption is that the histograms are the marginal distributions of a multivariate distribution
with independent components: i.e., given the $i-th$ description,
and the two $f_{i}(x)$ and $f_{i}(y)$ \emph{pdf}'s, and the joint
\emph{pdf} is expressed as $f_{i}(x,y)=f_{i}(x)\cdot f_{i}(y)$. In
this approach, interval data are considered as uniform \emph{pdf}'s,
and as a particular case of histogram with just one interval with
unitary weight. The regression method is based on the identification
of the covariance matrix that depends on the following basic statistics;
being $Y$ and $X_{j}$ the NPSV observed for $n$ individuals,
the means and the variances of each variable are computed according
to:

\begin{equation}
\bar{y}=\int\limits _{-\infty}^{+\infty}y\cdot f(y)dy=\frac{1}{n}\sum\limits _{i=1}^{n}\int\limits _{-\infty}^{+\infty}y\cdot f_{i}(y)dy=\frac{1}{n}\sum\limits _{i=1}^{n}\bar{y}_{i}
\end{equation}
\begin{equation}
s_{y}^{2}=\int\limits _{-\infty}^{+\infty}y^{2}\cdot f(y)dy-\bar{y}^{2}=\frac{1}{n}\sum\limits _{i=1}^{n}\int\limits _{-\infty}^{+\infty}y^{2}\cdot f_{i}(y)dy-\bar{y}^{2}
\end{equation}

and considering that
\begin{equation}
f(x,y)=\frac{1}{n}\sum\limits _{i=1}^{n}f_{i}(x,y)=\frac{1}{n}\sum\limits _{i=1}^{n}f_{i}(x)\cdot f_{i}(y)
\end{equation}

the covariance measure proposed by \citet{BER_GOUP_2000} is

\begin{equation}
s_{x,y}=\int\limits _{-\infty}^{+\infty}\int\limits _{-\infty}^{+\infty}x\cdot y\cdot f(x,y)dxdy-\bar{x}\cdot\bar{y}=\frac{1}{n}\sum\limits _{i=1}^{n}\bar{x}_{i}\bar{y}_{i}-\bar{x}\cdot\bar{y}.
\end{equation}

It is important to note that in general $s_{x}^{2}\neq s_{x,x}$.
\citet{billard2006symbolic} proposed a different way to compute $s_{x,y}$ when data are intervals, considering them as uniform
distributions as well as when data are histograms, considering them as weighted
intervals.

The proposal extends the linear regression model for standard data. 
%
%
%
%
%

Nevertheless, it is a method for predicting single values but not directly distributions
(this point is also considered by the authors). In general, it is
difficult to express the distribution of a linear combination of random
variables, in particular when the random variables are not identically
distributed. In general, an approximation of the distribution associated
to $\hat{y}$ can be obtained by means of a Montecarlo experiment.

On the other hand, if a set of strong conditions hold (the knowledge
of the cardinality of groups, the internal independence of the multivariate
distribution in each group), the model parameters have the same inferential
properties of the classic OLS linear regression estimation method.

\subsection{Wasserstein distance based models\label{sub:Wasserstein-distance-based}}

We have recalled that the \citet{billard2006symbolic} model is implicitly
founded on the modeling of the union of groups or concepts and
it is mainly based on the basic statistics proposed by \citet{BER_GOUP_2000}.
For example if there are two groups $G_{1}$ and $G_{2}$ of people
with the same cardinality, described by their income distributions
$f_{1}(INCOME)$ and $f_{2}(INCOME)$, all the basic statistics correspond
to the classic statistics calculated for the mixture of distributions.
From this point of view the basic statistics of the group $G_{1}\bigcup G_{2}$
are those calculated considering the \emph{pdf} of the variable \emph{INCOME} as:  $f(INCOME)=0.5f_{1}(INCOME)+0.5f_{2}(INCOME)$.
This situation arises frequently when the aim is to describe unions
of groups (for example, municipalities are grouped into cities).
In other cases, this approach can be inconsistent. For example, we
cannot know the cardinality of the groups or it makes no sense to know
the number of the elementary observations: if we take several pulse
rate measurements of two individuals $\omega_{1}$ and $\omega_{2}$
and we fit a distribution or a histogram $f_{1}(PulseRate)$ and $f_{2}(PulseRate)$
for each one of them, we may be interested to discover relations between
the two individuals by means of the comparison of their (probabilistic)
respective distributions, instead of considering a mixture of distributions (that is also a logical non
sense, two individuals cannot be fused into a super individual!).
\\
In this sense \citet{citeulike:7031164} proposed a different approach
based on the comparison of distributions by means of suitable dissimilarity
measures. \citet{citeulike:7031164} considered different kinds of
probabilistic metrics for histogram data and suggested that the same
results can be extended to data described by density functions (i.e.,
NPSD). Among the discussed metrics, the $\ell_2$ Wasserstein \citep{Wasse69}
distance permits to explain and interpret in an easy way the proximity
relations between two probability functions. Given two \emph{pdf}'s
$f(x)$and $g(x)$, with means $\bar{x}_{f}$ and $\bar{x}_{g}$,
finite standard deviations $s_{f}$ and $s_{g}$ it is possible to
associate respectively their \emph{cfd}'s $F(x)$ and $G(x)$. With
each \emph{cdf}'s it is associated their \emph{quantile functions
(qf)}, i.e. the inverse functions of the \emph{cdf}: $x_{f}(t)=F^{-1}(t)$
and $x_{g}(t)=G^{-1}(t)$. The $\ell_2$ Wasserstein distance is the following:

\begin{equation}
d_{W}(f,g)=\sqrt{\int\limits _{0}^{1}\left[x_{f}(t)-x_{g}(t)\right]^{2}dt}.\label{eq:WASS1}
\end{equation}

The $\ell_2$ Wasserstein distance is proposed for calculating the square errors
in the OLS problem. Given the matrix (\ref{eq:TAB_SYM}), we consider
the associated matrix $\mathbf{M}$ containing the corresponding quantile
functions:
\begin{equation}
\mathbf{M}=\left[\begin{array}{c|c}
\mathbf{Y} & \mathbf{X}\end{array}\right]=\left[\begin{array}{c|ccccc}
y_{1}(t) & x_{11}(t) & \cdots & x_{1j}(t) & \cdots & x_{1p}(t)\\
\cdots & \cdots & \cdots & \cdots & \cdots & \cdots\\
y_{i}(t) & x_{i1}(t) & \cdots & x_{ij}(t) & \cdots & x_{ipj}(t)\\
\cdots & \cdots & \cdots & \cdots & \cdots & \cdots\\
y_{n}(t) & x_{n1}(t) & \cdots & x_{nj}(t) & \cdots & x_{np}(t)
\end{array}\right];\label{eq:TAB_SYM-1}
\end{equation}

In this case, given a set of $p$ quantile functions for the $i-th$
individual, we look for a linear combination of $x_{ij}(t)$'s (for $j=1, \ldots,p$) which allows to predict
the $y_{i}(t)$'s (for $i=1, \ldots, n$) except for an error term $e_{i}(t)$.
It is worth noting that $e_{i}(t)$ is a residual function, not necessarily a quantile function
 The model to be fit is the following:

\begin{equation}
y_{i}(t)=\beta_{0}+\sum\limits _{j=1}^{p}\beta_{j}x_{ij}(t)+e_{i}(t),\quad\forall t\in [0,1]\label{eq:General_W_model}
\end{equation}

where the Sum of Squared Errors ($SSE$) to minimize for the solution
of the LS problem is defined according to the to the Squared $\ell_2$ Wasserstein distance
as follows:
\begin{equation}
SSE(\beta_0,\beta_1, \ldots, \beta_p)=\sum\limits _{i=1}^{n}d_{W}^{2}\left(y_{i}(t),\left[\beta_{0}+\sum\limits _{j=1}^{p}\beta_{j}x_{ij}(t)\right]\right)=\sum\limits _{i=1}^{n}\int\limits_0^1 \left[e_{i}(t)\right]^{2}dt.\label{eq:SS_stand}
\end{equation}
A problem arises for the linear combination of quantile functions:
 only if $\beta_{j}\geq0\:(j=1,\ldots,p)$ it is assured
that $y_{i}(t)$ is a quantile function (i.e. a not decreasing function).
In order to overcome this problem, \citet{dias_brito_reg_2011}
proposed a novel method for the regression of histogram valued data
based on the Wasserstein distance between quantile functions. They
proposed to expand the matrix $\mathbf{M}$ adding also the quantile
functions of the \emph{symmetric distributions} of the explicative symbolic
variables. Given $f_{i}(x_{j})$ (and the respective quantile function $x_{ij}(t)$),
the corresponding symmetric distribution $\tilde{f}_{i}(x_{j})$ (and
its quantile function $\tilde{x}_{ij}(t)$) is obtained by multiplying
the support of {$f_{i}(x_{j})$} by $-1$,
such that the (integral of) the sum of the two quantile functions is equal to zero (${\int\limits _{0}^{1}\left[x_{ij}(t)+\tilde{x}_{ij}(t)\right]dt=0}$).

The model is the following:

\begin{equation}\label{eq:Brito}
y_{i}(t)=\beta_{0}+\sum\limits _{j=1}^{p}\beta_{j}x_{ij}(t)+\sum\limits _{j=1}^{p}\tilde{\beta}_{j}\tilde{x}_{ij}(t)+e_{i}(t),
\end{equation}

and the estimation of the parameters is obtained optimizing the following
constrained OLS problem:

\begin{eqnarray*}
\underset{(\beta_{0},\beta_{j},\tilde{\beta}_{j})}{argmin} \, SSE & = & \sum\limits _{i=1}^{n}d_{W}^{2}\left(y_{i}(t),\left[\beta_{0}+\sum\limits _{j=1}^{p}\beta_{j}x_{ij}(t)+\sum\limits _{j=1}^{p}\tilde{\beta}_{j}\tilde{x}_{ij}(t)\right]\right)\\
s.a &  & \beta_{j},\tilde{\beta}_{j}\geq0
\end{eqnarray*}

\subsection{The novel LS method for multiple regression\label{sub:The-Irpino-and}}

The negative value of the parameter $\beta_{j}$ in model (\ref{eq:General_W_model})
is in general not acceptable when dealing with quantile functions.
In order to overcome this inconvenient, starting from a particular
decomposition of the Wasserstein \citet{VE_IRP_2010_OLS} presented
a new formulation of the problem in the simple regression model, that is when only one variable \emph{X} affects the predicted variable \emph{Y}.

The introduction of the new model can be done according to some preliminary
considerations about the properties of $\ell_2$ Wasserstein distance decomposition.
Given $f(x)$ and $g(x)$ two NPSD and being $\bar{x}_{f}$ and $\bar{x}_{g}$ the respective means,
$x_{f}^{c}(t)$ and $x_{g}^{c}(t)$  the respective centred quantiles functions (above defined in \S2 \emph{Notation and definitions})
, \citet{Cuesta-Albertos:1997:OTP:255396.255406} showed that the
(Squared) $\ell_2$ Wasserstein distance can be rewritten as
\begin{equation}
d_{W}^{2}(f,g)=\left(\bar{x}_{f}-\bar{x}_{g}\right)^{2}+{\int\limits _{0}^{1}\left[x_{f}^{c}(t)-x_{g}^{c}(t)\right]^{2}dt}.\label{eq:WASS2}
\end{equation}
This property allows to consider the squared distance as the sum of
two components, the first related to the location of NPSD and the
second related to their variability structure. \citet{DBLP:conf/f-egc/2007}
improved such decomposition, showing that the $d_{W}^{2}$
can be finally decomposed into three quantities:
\begin{equation}
d_{W}^{2}(f,g)=\left(\bar{x}_{f}-\bar{x}_{g}\right)^{2}+\left(s_{f}-s_{g}\right)^{2}+2s_{f}s_{g}\left(1-\rho(x_{f},x_{g})\right)\label{eq:WASS3comp}
\end{equation}

where $\rho(x_{f},x_{g})$ is a correlation coefficient about the
quantile functions, i.e.:
\begin{equation}
\rho(x_{f},x_{g})=\frac{\int_{0}^{1}x_{f}(t)\cdot x_{g}(t)dt-\bar{x}_{f}\cdot\bar{x}_{g}}{s_{f}\cdot s_{g}}.\label{eq:rhoQQ}
\end{equation}

\citet{Irpino2006} showed some computational aspects relating to
$\rho$ when data are histograms and  they showed that $\rho$ is computed
in a linear time with the total number of bins of the histograms.
The Equation \ref{eq:rhoQQ} allows to consider the (inner) product between two
\emph{qf}'s, in a functional mood, as follows:
\begin{equation}
\left\langle x_{f}(t),x_{g}(t)\right\rangle =\int_{0}^{1}x_{f}(t)\cdot x_{g}(t)dt=\rho(x_{f},x_{g})\cdot s_{f}\cdot s_{g}+\bar{x}_{f}\cdot\bar{x}_{g}.\label{eq:prodQuantil}
\end{equation}

Thus, given two vectors of quantile functions $\mathbf{x}=[x_{i}(t)]_{n\times1}$and
$\mathbf{y}=[y_{i}(t)]_{n\times1}$, we define the scalar product
of two vectors of NPSD as:
\begin{equation}
\mathbf{x^{T}y}=\sum\limits _{i=1}^{n}\left\langle x_i{t}(t),y_{i}(t)\right\rangle=\sum\limits _{i=1}^{n}\left[\rho(x_{i},y_{i})\cdot s_{x_{i}}\cdot s_{y_{i}}+\bar{x}_{i}\cdot\bar{y}_{i}\right]. \label{eq:prod_vect_quant}
\end{equation}
It is important noting that Eq. (\ref{eq:prod_vect_quant}) allows to express the product of two matrices of quantile functions, and it corresponds exactly to the product of two classic matrices of punctual data (in that case $s_{x_{i}}=0$ and $s_{y_{i}}=0$). The inner product in Eq. (\ref{eq:prodQuantil})is the direct consequence of having a space of quantile functions equipped with the $\ell_2$ Wasserstein norm. Thus, we can do classic matrix operations on matrices of quantile functions.

If we consider that $x_{ij}^{c}(t)=x_{ij}(t)-\bar{x}_{ij}$, each
element of $\mathbf{X}$ can be rewritten as $x_{ij}(t)=x_{ij}^{c}(t)+\bar{x}_{ij}$.
The same is valid for vector ${\bf Y}$. Matrix ${\bf M}$ is
transformed as\footnote{Note that $\left[\mathbf{Y}|\mathbf{X}\right]$ denote a block matrix of two elements along the columns.}:

\begin{equation}
\mathbf{M}=\left[\begin{array}{c|c}
\mathbf{Y} & \bar{\mathbf{X}}+\mathbf{X^{c}}\end{array}\right]=\left[\begin{array}{c|c}
\mathbf{Y} & \mathbf{\bar{X}}\end{array}\right]+\left[\begin{array}{c|c}
\mathbf{Y} & \mathbf{X^{c}}\end{array}\right]\label{eq:TAB_SYM-1-1}
\end{equation}
 where\footnote{Please, note that the sum of  quantile function $x(t)$ and a scalar value $\alpha$ is equal to $[\alpha+x](t)=\alpha+x(t)\quad\forall t\in[0,1]$. While the sum of two quantile functions $x(t)$ and $y(t)$ is $[x+y](t)=x(t)+y(t)\quad\forall t\in[0,1]$.}
$\mathbf{\bar{X}}=\left[\bar{x}_{ij}\right]_{n\times p}$ is the matrix
of the means of the $f_{i}(x_{j})$, $\mathbf{X^{c}}=\left[x_{ij}^{c}(t)\right]_{n\times p}$
is the matrix of the centred quantile functions of $f_{i}(x_{j})$'s.

We assume that each quantile function $y_{i}(t)$ can be expressed
as a linear combination of the means $\bar{x}_{ij}$ and of the centred quantile functions $x_{ij}^{c}(t)$
plus an error term $e_{i}(t)$ (which is a function) as follows:

\begin{equation}
y_{i}(t)=\beta_{0}+\sum\limits _{j=1}^{p}\beta_{j}\bar{x}_{ij}+\sum\limits _{j=1}^{p}\gamma_{j}x_{ij}^{c}(t)+e_i(t), \quad \forall t \in [0,1] \label{eq:model_no_mat}
\end{equation}

If we consider the matrix ${\bf \bar{X}_{+}=[{\bf 1|{\bf \bar{X}]}}}$,
we can rewrite the model in (\ref{eq:model_no_mat}) using the matrix
notation as follows:

\begin{equation}
\mathbf{Y}={\bf \bar{X}_{+}B+{\bf X^c\mathbf{\Gamma}+{\bf{e}}}}\label{eq:multimodel}
\end{equation}

In order to estimate the parameters, we define the Sum of Square Errors
function (SSE) like in the LS method, using the Wasserstein $\ell_2$ measure:

\begin{equation}
\begin{array}{l}
SSE(\beta_0,\beta_j,\gamma_j)=\sum\limits _{i=1}^{n}\int\limits _{0}^{1}e_{i}^{2}(t)dt=\\
=\sum\limits _{i=1}^{n}\int\limits _{0}^{1}\left[y_{i}(t)-\left(\beta_{0}+\sum\limits _{j=1}^{p}\beta_{j}\bar{x}_{ij}+\sum\limits _{j=1}^{p}\gamma_{j}x_{ij}^{c}(t)\right)\right]^{2}\\

\end{array}
\end{equation}
 in matrix form:
\begin{equation}
SSE\mathbf{(B,\Gamma)=e^{T}e}=\mathbf{\left[Y-\bar{X}_{+}B-X^{c}\Gamma\right]^{T}\left[Y-\bar{X}_{+}B-X^{c}\Gamma\right]}\label{eq:SSgen}
\end{equation}

Considering the equation (\ref{eq:multimodel}) induced from the
$\ell_2$ Wasserstein metric, we have ${\bf \bar{X}}_{+}^{T}{\bf X}^{{\bf c}}=\mathbf{0}_{(p+1)\times p}$,
${\bf \bar{X}}_{+}^{T}\mathbf{Y}={\bf \bar{X}}_{+}^{T}\mathbf{\bar{Y}}$
and ${\bf X}^{{\bf c}}{}^{T}{\bf Y}={\bf X}^{{\bf c}}{}^{T}{\bf Y}^{\mathbf{c}}$\footnote{Some algebraic details are in appendix\ref{sec:OLS-solution-details}}.
Then, SSE in equation (\ref{eq:SSgen}) can be decomposed into two positive quantities as
follows:

\begin{equation}
SSE(\mathbf{B,\Gamma})=SSE(\mathbf{B})+SSE(\mathbf{\Gamma})=
\bar{\mathbf{e}}^{T}\bar{\mathbf{e}}+\mathbf{\left(e^{c}\right)^{T}e^{c}}\label{eq:SS_two}
\end{equation}

where:
\begin{eqnarray}
\mathbf{\bar{e}} & = & \mathbf{\bar{Y}-\bar{X}_{+}B}\label{eq:mod_me1}\\
\mathbf{e^{c}} & = & \mathbf{Y^{c} - X^{c}\Gamma}\label{eq:mod_centr1}
\end{eqnarray}

with ${\bf \bar{e}=[\bar{e}_{i}]_{n\times1}}$ a vector of real numbers.

We may express the single minimization problem as the minimization of two independent functions:
the first one related to the means of the predictor quantile functions $\bar{x}_{ij}$'s in $\bf \bar{X}_{+}$, and the second one related to the variability of the centered quantile distributions $x^c_{ij}(t)$'s in $ \bf X^{c}$. Then two models are independently estimated:

\begin{eqnarray}
\mathbf{\bar{Y}} & = & \mathbf{\bar{X}_{+}B+\bar{e}}\label{eq:mod_me2}\\
\mathbf{Y^{c}} & = & \mathbf{X^{c}\Gamma+e^{c}}\label{eq:mod_centr2}\label{eq:GAMMAS}
\end{eqnarray}

The first can be solved as classic LS problem for the estimation of $\mathbf{B}$:
\begin{equation}
\underset{\mathbf{B}}{argmin}\, SSE(\mathbf{B})=\bar{\mathbf{e}}^{T}\bar{\mathbf{e}}=\left[\mathbf{\bar{Y}-\bar{X}_{+}B}\right]^{T}\left[\mathbf{\bar{Y}-\bar{X}_{+}B}\right]\label{eq:ols_betas}
\end{equation}


The second model in Eq. (\ref{eq:GAMMAS}) is solved using the NNLS (Non
Negative Least Squares) algorithm proposed by \citet{Law_hans_74_NNLS},
modified with the introduction of the product between quantile functions
(Eq. \ref{eq:prodQuantil}) in the classic matrix computations:
\begin{eqnarray}
\underset{\Gamma}{argmin}\, SSE(\mathbf{\Gamma})& = & \mathbf{\left(e^{c}\right)^{T}e^{c}} = \left[\mathbf{\mathbf{Y^{c}-X^{c}\Gamma}}\right]^{T}\left[\mathbf{Y^{c}-X^{c}\Gamma}\right]\label{eq:ols_gamma}\\
s.a. &  & \gamma_{j}\geq0\: j=1,\ldots,p.\nonumber
\end{eqnarray}

The estimated LS parameters are:

\begin{equation}
{\bf \hat{B}}=\left({\bf \bar{X}}_{+}^{T}\mathbf{\bar{X}}\right)^{-1}{\bf \bar{X}}_{+}^{T}{\bf \bar{Y}}\label{eq:slopes}
\end{equation}
\begin{equation}
{\bf \hat{\Gamma}}=\left({\bf X}^{{\bf c}}{}^{T}{\bf X}^{{\bf c}}\right)^{-1}{\bf X}^{{\bf c}}{}^{T}{\bf Y}^{{\bf c}}.\label{eq:GAMMA}
\end{equation}

Therefore, a \emph{qf} $\hat{y}_{i}(t)$ is predicted by the  estimated linear model according to the estimated parameters:
\begin{equation}
\hat{y}_{i}(t)=\hat{\bar{y}}_{i}+\hat{y}_{i}^{c}(t)=\hat{\beta}_{0}+\sum\limits _{j=1}^{p}\hat{\beta}_{j}\bar{x}_{ij}+\sum\limits _{j=1}^{p}\hat{\gamma}_{j}x_{ij}^{c}(t).\label{eq:prediction}
\end{equation}


\section{The goodness of fit evaluation\label{sec:The-goodness-of}}

Considering the nature of the data, the evaluation of the goodness
of fit of the model is not straightforward like for the classic linear
regression models. Here we present three indices that can be used
for evaluating the goodness of fit of a regression model on NPSD.
The first is the $\Omega$ measure proposed by \citet{dias_brito_reg_2011},
the second is the $Pseudo-R^{2}$ proposed by \citet{VE_IRP_2010_OLS}
and the last is the classic square root of the mean sum of squares
expressed using the $\ell_2$ Wasserstein distance (we denote is as $RMSE_{W}$).
The three measures are detailed in the following.
\begin{description}
\item [{$\mathbf{\Omega}$ \cite{dias_brito_reg_2011}}] The proposed measure
is the ratio
\begin{equation}
\Omega=\frac{\sum\limits _{i=1}^{n}d_{W}^{2}\left(\hat{y}_{i}(t),\bar{y}\right)}{\sum\limits _{i=1}^{n}d_{W}^{2}\left(y_{i}(t),\bar{y}\right)}
\end{equation}
\\
that varies from zero to 1.
\item [{$\mathbf{Pseudo-R^{2}}$}] It is a measure proposed by \citet{VE_IRP_2010_OLS}
for the simple linear regression model. \citet{VER_IR_CP2008} proved that
the Wasserstein distance can be used for the definition of a sum of
squared deviation as follows:
\begin{equation}
SSY=n\cdot s_{y}^{2}=\sum\limits _{i=1}^{n}d_{W}^{2}\left(y_{i}(t),\bar{y}(t)\right)=\sum\limits _{i=1}^{n}\int\limits _{0}^{1}\left[y_{i}(t)-\bar{y}(t)\right]^{2}dt.\label{eq:dev_irp}
\end{equation}

where $\bar{y}(t)$ is the \emph{baricenter} of NPSD $y_i(t)$'s as defined at the end of section 2.

A common tool for the evaluation of the goodness of fit of the model is the well
known \emph{coefficient of determination} $R^{2}$ ($R^{2}=\frac{SSR}{SSY}$ or $R^{2}=1-\frac{SSE}{SSY}$).

The computation of this coefficient is based on the partitions of the total variation in the dependent variable, denoted SST, into two parts: the part explained by the estimated regression equation, denoted SSR, and the part that measures the unexplained variation, SSE, referred to as the residual sum of squares:
\begin{equation}
SSY=SSE+SSR
\end{equation}

In our case, in general, the equality does not hold,
and we prove%
\footnote{See  \ref{sec:Sum-of-square}.%
} that the decomposition of $SSY$ is the following:
\begin{equation}
\begin{array}{l}
SSY=\underbrace{\sum\limits _{i=1}^{n}\int\limits _{0}^{1}\left[\hat{y}_{i}(t)-y_{i}(t)\right]^{2}dt}_{SSE}+\underbrace{\sum\limits _{i=1}^{n}\int\limits _{0}^{1}\left[\bar{y}(t)-\hat{y}_{i}(t)\right]^{2}dt}_{SSR}+\\
\underbrace{-2\left[n\cdot\left(\sigma_{\bar{y}}^{2}-\sum\limits _{j=1}^{p}\gamma_{j}r_{\bar{y}\bar{x}_{j}}\sigma_{\bar{y}}\sigma_{\bar{x}_{j}}\right)+{\bf \boldsymbol{\Gamma}}\nabla SSE({\bf \boldsymbol{\Gamma}})\right]}_{Bias}.
\end{array}\label{eq:SSY_M}
\end{equation}

The $\emph{Bias}$ term in eq. (\ref{eq:SSY_M}) reflects the impossibility of the linear transformation of $\bar{x}(t)$ of reflecting the variability
structure of $\bar{y}(t)$. In general, this term goes to zero when
NPSD have the same shape (i.e., from the third ones forward, the standardized
moments of the histograms are equal) and the standard deviations of
$f_{ij}(x)$'s are proportional to the standard deviations of the
$f_{i}(y)$'s.
Further, the term ${\bf \boldsymbol{\Gamma}}\nabla SSE({\bf \boldsymbol{\Gamma}})$
indicates the difference between the classic OLS solutions (that may lead to negative $\gamma$'s) and  solution obtained using the
NNLS. \\
In this case, the classic $R^{2}=1-\frac{SSE}{SSY}$ statistic can
be less than zero or greater than 1. In order to obtain a measure of goodness of fit
that does not suffer of the described drawback, we propose to adopt
the following general index, that takes values in  $[0,1]$:
\begin{equation}
PseudoR^{2}=\min\left[\max\left[0;1-\frac{SSE}{SSY}\right];1\right].
\end{equation}

\item [{RMSE}]
The Root Mean Square Error is generally used as measure
of goodness of fit. Choosing an appropriate measure for computing
the distance between NPSD's, we may compute the RMSE. In this paper,
having used the Wasserstein distance, we propose the following measure
for the RMSE:
\begin{equation}
RMSE_{W}=\sqrt{\frac{\sum\limits _{i=1}^{n}\int\limits _{0}^{1}\left(\hat{y}_{i}(t)-y_{i}(t)\right)^{2}dt}{n}}=\sqrt{\frac{SSE}{n}}.
\end{equation}

\end{description}

\section{Application on real data}\label{sec:apply}

To illustrate the proposed method we choose some examples presented
in the literature on clinic data and a new climatic dataset. Especially, we make use only of dataset of NPSD described
by histograms usually arisen as summaries of large amount of data.

The first dataset is presented by \citet[Chap.6,  Table 6.8]{billard2006symbolic}
and also presented as application in \citet{dias_brito_reg_2011}.
In the dataset presented in table \ref{TAB:Blood_dataset} there are
the Hematocrit (Y) histogram NPSD and the Hemoglobin (X) histogram
NPSD observed for 10 units.

\begin{table}[th]
\centering{}%
\begin{tabular}{lll}
\hline
Units & Y: Hematocrit  & X: Hemoglobin\\
\hline
1 & {\small \{{[}33.29;37.57{[}, }\textbf{{\small 0.6}}{\small{} ;
{[}37,52; 39.61{]}, }\textbf{\small 0.4}{\small \}} & {\small \{{[}11.54; 12.19{[} , }\textbf{\small 0.4}{\small ; {[}12.19;
12.8{]} ,}\textbf{\small{} 0.6}{\small \}}\\
2 & {\small \{{[}36.69; 39.11{[} , }\textbf{\small 0.3}{\small ; {[}39.11;
45.12{]} , }\textbf{\small 0.7}{\small \} } & {\small \{{[}12.07; 13.32{[} , }\textbf{\small 0.5}{\small ; {[}13.32;
14.17{]} , }\textbf{\small 0.5}{\small \}}\\
3 & {\small \{{[}36.69; 42.64{[} , }\textbf{\small 0.5}{\small ; {[}42.64;
48.68{]} , }\textbf{\small 0.5}{\small \} } & {\small \{{[}12.38; 14.2{[} , }\textbf{\small 0.3}{\small ; {[}14.2;
16.16{]} , }\textbf{\small 0.7}{\small \}}\\
4 & {\small \{{[}36.38; 40.87{[} , }\textbf{\small 0.4}{\small ; {[}40.87;
47.41{]} , }\textbf{\small 0.6}{\small \} } & {\small \{{[}12.38; 14.26{[} , }\textbf{\small 0.5}{\small ; {[}14.26;
15.29{]} , }\textbf{\small 0.5}{\small \}}\\
5 & {\small \{{[}39.19; 50.86{]} , }\textbf{\small 1}{\small \} } & {\small \{{[}13.58; 14.28{[} , }\textbf{\small 0.3}{\small ; {[}14.28;
16.24{]} , }\textbf{\small 0.7}{\small \}}\\
6 & {\small \{{[}39.7; 44.32{[} , }\textbf{\small 0.4}{\small ; {[}44.32;
47.24{]} ,}\textbf{\small{} 0.6}{\small \} } & {\small \{{[}13.81; 14.5{[} ,}\textbf{\small{} 0.4}{\small ; {[}14.5;
15.2{]} , }\textbf{\small 0.6}{\small \}}\\
7 & {\small \{{[}41.56; 46.65{[} , }\textbf{\small 0.6}{\small ; {[}46.65;
48.81{]} , }\textbf{\small 0.4}{\small \} } & {\small \{{[}14.34; 14.81{[} ,}\textbf{\small{} 0.5}{\small ; {[}14.81;
15.55{]} ,}\textbf{\small{} 0.5}{\small \}}\\
8 & {\small \{{[}38.4; 42.93{[} , }\textbf{\small 0.7}{\small ; {[}42.93;
45.22{]} , }\textbf{\small 0.3}{\small \} } & {\small \{{[}13.27; 14.0{[} ,}\textbf{\small{} 0.6}{\small ; {[}14.0;
14.6{]} ,}\textbf{\small{} 0.4}{\small \} }\\
9 & {\small \{{[}28.83; 35.55{[} , }\textbf{\small 0.5}{\small ; {[}35.55;
41.98{]} , }\textbf{\small 0.5}{\small \} } & {\small \{{[}9.92; 11.98{[} , }\textbf{\small 0.4}{\small ; {[}11.98;
13.8{]} , }\textbf{\small 0.6}{\small \}}\\
10 & {\small \{{[}44.48; 52.53{]} , }\textbf{\small 1}{\small \} } & {\small \{{[}15.37; 15.78{[} , }\textbf{\small 0.3}{\small ; {[}15.78;
16.75{]} , }\textbf{\small 0.7}{\small \}}\\
\hline
\end{tabular} \caption{Blood dataset: 10 units described by two histogram-valued variables.}\label{TAB:Blood_dataset}
\end{table}

\begin{figure}
\begin{centering}
\includegraphics[scale=0.7]{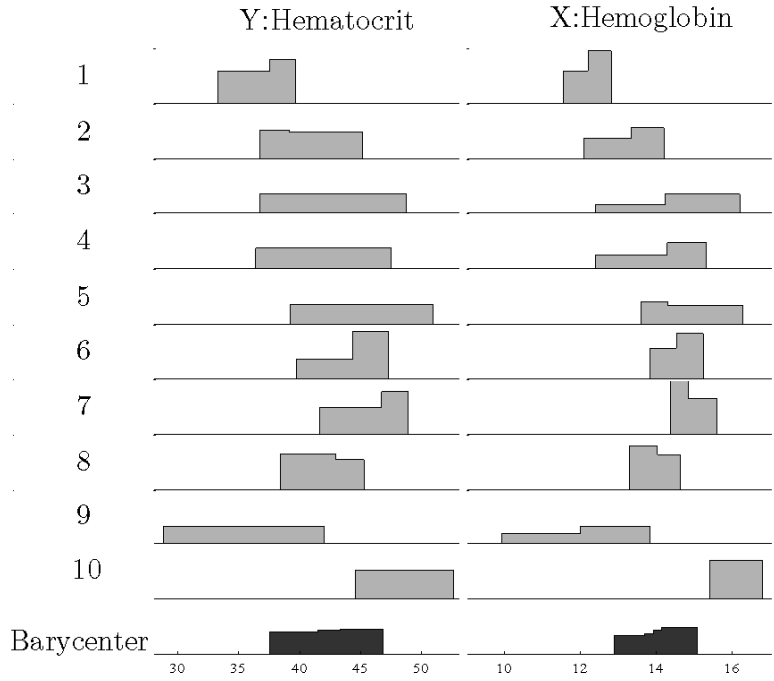}
\end{centering}
\caption{Blood dataset. Histogram representation, in the last row are represented
the barycenter histograms.}\label{Fig: Blood_data}
\end{figure}

Fig. \ref{Fig: Blood_data} shows the graphical representation
of table \ref{TAB:Blood_dataset} and the graphic representation of
the means (barycenter) NPSD of each histogram variable, according to the barycenter
of histogram variable as presented in \citep{VER_IR_CP2008}.\\

Table \ref{TAB:Blood_summary} specifies the main summary statistics
for the two histogram variables using the \citet{billard2006symbolic}
set of summary statistics and those proposed by \citet{VER_IR_CP2008}.
For the barycenters the mean and the standard deviation  are only reported.
Obviously, it is possible to report also the other moments of the barycenters,
but for the sake of brevity we prefer to leave to the reader further
considerations looking directly to their graphical representations
in fig. \ref{Fig: Blood_data}.\\

\begin{table}
\centering{}%
\begin{tabular}{lcc}
\hline
 & Y:Hemoglobin & X: Hematocrit\\
\hline
n & \multicolumn{2}{c}{10}\\
Mean (BD) & 42.26 & 14.05\\
Barycenter mean(VI) & 42.26 & 14.05\\
Barycenter std (VI) & 2.660 & 0.622\\
Standard deviation (BD) & 4.658 & 1.355\\
Standard deviation (VI) & 3.824 & 1.204\\
Correlation (BD) & \multicolumn{2}{c}{0.903}\\
Correlation (VI) & \multicolumn{2}{c}{0.979}\\
\hline
\end{tabular}\caption{Blood dataset: summary statistics. BD refers to the approach of \citep{billard2006symbolic}, while VI refers to the approach of \citep{VER_IR_CP2008}}\label{TAB:Blood_summary}
\end{table}

In Tabs. \ref{TAB:Blood_Bill_Did}, \ref{TAB:Blood_Dia_Brito} and \ref{TAB:Blood_Irp_Ver}, the rows labeled with \emph{Observed} reports the OLS estimates of the parameters of the
regression models using the formulation of \citet{billard2006symbolic}
as described in \ref{sub:The-Billard-Diday-model}, the formulation
of \citet{dias_brito_reg_2011} as presented in \ref{sub:Wasserstein-distance-based}
and the novel formulation as proposed in \ref{sub:The-Irpino-and}.

\begin{table}
\begin{centering}
\begin{tabular}{lccccc}
\multicolumn{6}{c}{Billard-Diday Model: $\hat{y}_{i}=\hat{\beta_{0}}+\hat{\beta}_{1}x_{i}$}\\
\cline{2-6}
 & \multicolumn{2}{c}{Model parameters} & \multicolumn{3}{c}{Goodness of fit }\\
\cline{2-6}
 &
$\hat{\beta_{0}}$
 &
$\hat{\beta}_{1}$
 & $\Omega$ & {\small Pseudo}$R^{2}$ & $RMSE_{W}$\\
\cline{2-6}
Observed & {\small -1.336} & {\small 3.103} & \emph{\footnotesize 0.811} & \emph{\footnotesize 0.919} & \emph{\footnotesize 1.085}\\
 & \multicolumn{5}{c}{{\small Bootstrap estimates}}\\
Mean & {\small 2.224} & {\small 2.851} & \emph{\footnotesize 0.692} & \emph{\footnotesize 0.812} & \emph{\footnotesize 1.221}\\
Bias & {\small 3.561} & {\small -0.253} &  &  & \\
SE & {\small 6.592} & {\small 0.461} &  &  & \\
2.5\% & {\small -4.528} & {\small 1.609} & \emph{\footnotesize 0.197} & \emph{\footnotesize 0.155} & \emph{\footnotesize 0.857}\\
97.5\% & {\small 20.243} & {\small 3.318} & \emph{\footnotesize 0.888} & \emph{\footnotesize 0.953} & \emph{\footnotesize 2.167}\\
\cline{2-6}
\end{tabular}
\end{centering}
\caption{Blood dataset: Billard-Diday model parameters estimated on the full
dataset and bootstrapping the dataset.}\label{TAB:Blood_Bill_Did}
\end{table}
In order to compare the three models, we computed the goodness of
fit measures presented in \ref{sec:The-goodness-of}. We remark that
Billard-Diday model does not allow directly to predict a distribution
function or a quantile function associated with $\hat{f}_{i}(y)$,
thus, for computing the goodness of fit indices of the Billard-Diday
model, we performed a Montecarlo experiment in order to estimate
the predicted $\hat{f}_{i}(y)$ distribution.\\
\begin{table}
\begin{centering}
\begin{tabular}{ccccccc}
\multicolumn{7}{c}{Dias-Brito Model: $\hat{y}_{i}(t)=\hat{\beta_{0}}+\hat{\beta}_{1}\cdot x_{i}(t)+\hat{\tilde{\beta}}_{1}\cdot\tilde{x}_{i}(t)$}\\
\cline{2-7}
 & \multicolumn{3}{c}{Model parameters} & \multicolumn{3}{c}{Goodness of fit }\\
\cline{2-7}
 & $\hat{\beta_{0}}$  & $\hat{\beta}_{1}$  & $\hat{\tilde{\beta}}_{1}$ & $\Omega$ & {\small Pseudo}$R^{2}$ & $RMSE_{W}$\\
\cline{2-7}
Observed & -1.953 & 3.560 & 0.413 & 0.963 & 0.945 & 0.895\\
 & \multicolumn{6}{c}{Bootstrap estimates}\\
Mean & -1.657
 & 3.574 & 0.448 & 0.963 & 0.935 & 0.829
\\
Bias & 0.296 & 0.014 &  0.035 &  &  & \\
SE & 2.862 & 0.165 &  0.164 &  &  & \\
2.5\% & -5.848 & 3.255 & 0.217 & 0.928 & 0.823 & 0.528\\
97.5\% & 5.037 & 3.931 & 0.848 & 0.986 & 0.981 & 1.058
\\
\cline{2-7}
\end{tabular}
\end{centering}
\caption{Blood dataset: Dias-Brito model parameters estimated on the full dataset
and bootstrapping the dataset.}\label{TAB:Blood_Dia_Brito}
\end{table}
In order to calculate the confidence interval of the parameters of
the three models, and considering the complexity of establishing manageable
probabilistic hypotheses on the error functions $e_i(t)$ like in the classic linear regression estimation
problem, we have performed the bootstrap \citep{EFRON1993} estimates
of the parameters of the models by constructing $1,000$ resamples
of the observed dataset (and of equal size to the observed dataset),
each of which is obtained by random sampling with replacement from
the original dataset. The confidence interval of each parameter is
calculated using the percentile method, i.e. by using the 2.5 and
the 97.5 percentiles of the bootstrap distribution as the limits of
the 95\% confidence interval for each parameter.
\begin{table}
\begin{centering}
\begin{tabular}{ccccccc}
\multicolumn{7}{c}{Irpino-Verde model:{$\hat{y}_{i}(t)=\hat{\beta_{0}}+\hat{\beta}_{1}\cdot\bar{x}_{i}+\hat{\gamma}_{1}\cdot x_{i}^{c}(t)$}}\\
\cline{2-7}
 & \multicolumn{3}{c}{Model parameters} & \multicolumn{3}{c}{Goodness of fit }\\
\cline{2-7}
 & $\hat{\beta_{0}}$ & $\hat{\beta}_{1}$ & $\hat{\gamma}_{1}$ & $\Omega$ & {\small Pseudo}$R^{2}$ & $RMSE_{W}$\\
\cline{2-7}
Observed & -2.157 & 3.161 & 3.918 & \emph{\small 0.961} & \emph{\small 0.943} & \emph{\small 0.914}\\
 & \multicolumn{6}{c}{Bootstrap estimates}\\
Mean & -1.928 & 3.146 & 3.969 & \emph{\small 0.961} & \emph{\small 0.931 } & \emph{\small 0.850 }\\
Bias & 0.229 & -0.016 & 0.051 &  &  & \\
SE & 2.833 & 0.199 & 0.269 &  &  & \\
2.5\% & -6.348 & 2.688 & 3.602 & \emph{\small 0.924} & \emph{\small 0.812 } & \emph{\small 0.547 }\\
97.5\% & 4.644 & 3.462 & 4.710 & \emph{\small 0.985} & \emph{\small 0.980 } & \emph{\small 1.076 }\\
\cline{2-7}
\end{tabular}
\par\end{centering}
\caption{Blood dataset: Irpino-Verde model parameters estimated on the full
dataset and bootstrapping the dataset.}\label{TAB:Blood_Irp_Ver}
\end{table}

As usual, the point
estimates is the mean bootstrap value. The main results about the
parameters and the goodness of fit indices are shown in Tables
\ref{TAB:Blood_Bill_Did},\ref{TAB:Blood_Dia_Brito} and \ref{TAB:Blood_Irp_Ver}.

The results show that the Dias-Brito and the novel proposed method
fit better than the Billard-Diday model, and that the Dias-Brito and
the proposed method have negligible differences considering the goodness
of fit. The difference is about the interpretation of the regression
parameters. The parameters of the Billard-Diday model are interpretable
as for classic regression but the dependent
predicted distributions cannot be easily described. The Dias-Brito model parameters show that
a unitary (positive) variation of all the quantiles of the Hematocrit
induces a variation of 3.574 of the Hemoglobin quantiles, while a
unitary increase of the {}``symmetric'' Hematocrit quantiles induces
an increase of the Hemoglobin quantiles equal to 0.448, this can
be a bit confusing considering that there is a positive effect both
on the original dependent variable and on its symmetric version.

The interpretation of the Irpino-Verde model is different as it takes into consideration the
two components of the NPSD: the variability of the means and the variability
of the centered NPSD (the variability of the distributions variability).
The estimated model enounces that a unitary variation of the mean
of the Hematocrit induces a variation of 3.146 in the mean of the
Hemoglobin, while an increasing of one in the variability of the Hematocrit
produce an increase of 3.969, in average, of the variability of the Hematocrit.\\

\begin{figure}
\begin{centering}
\includegraphics[width=0.9\textwidth]{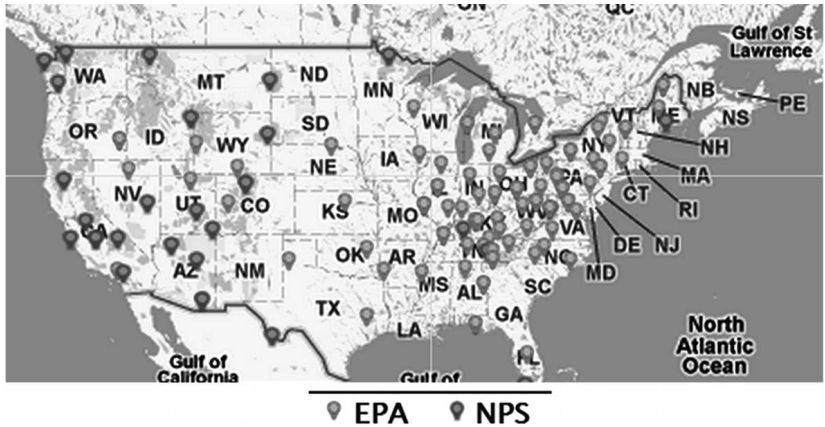}
\end{centering}
\caption{Ozone dataset. EPA and NPS monitored sites.}\label{FIG:EPA}
\end{figure}

The second dataset derives from the Clean Air Status and Trends Network
(CASTNET)%
\footnote{http://java.epa.gov/castnet/%
}, an air quality monitoring network of United States designed to provide
data to assess trends in air quality, atmospheric deposition, and
ecological effects due to changes in air pollutant emissions. In particular,
we have chosen to select data on the Ozone concentration in 78 USA
sites among those depicted in Fig. \ref{FIG:EPA} for which the
monitored data was complete (i.e. without missing values for each
of the selected characteristics).

Ozone is a gas that can cause respiratory diseases. In the literature
there exists studies that relate the Ozone concentration level to
the Temperature, the Wind speed and the Solar radiation (see for example
\citep{Duenas2002}).

\begin{figure}
\includegraphics[width=0.9\textwidth]{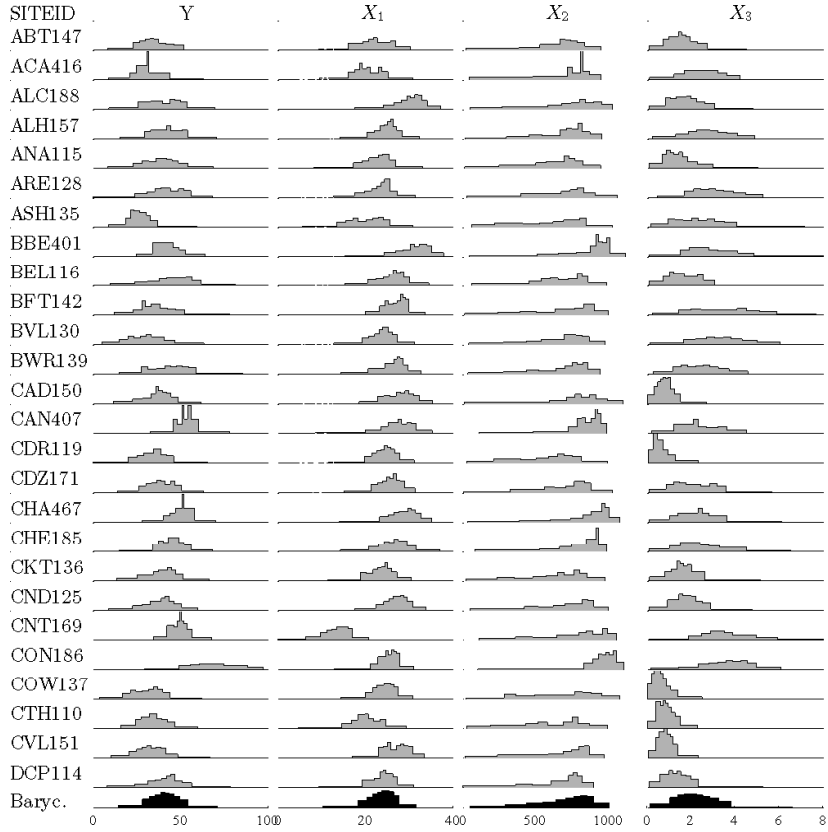}
\caption{Ozone dataset: monitored sites from 1 to 26.}\label{FIG:OZONE1}
\end{figure}

\begin{figure}
\includegraphics[width=0.9\textwidth]{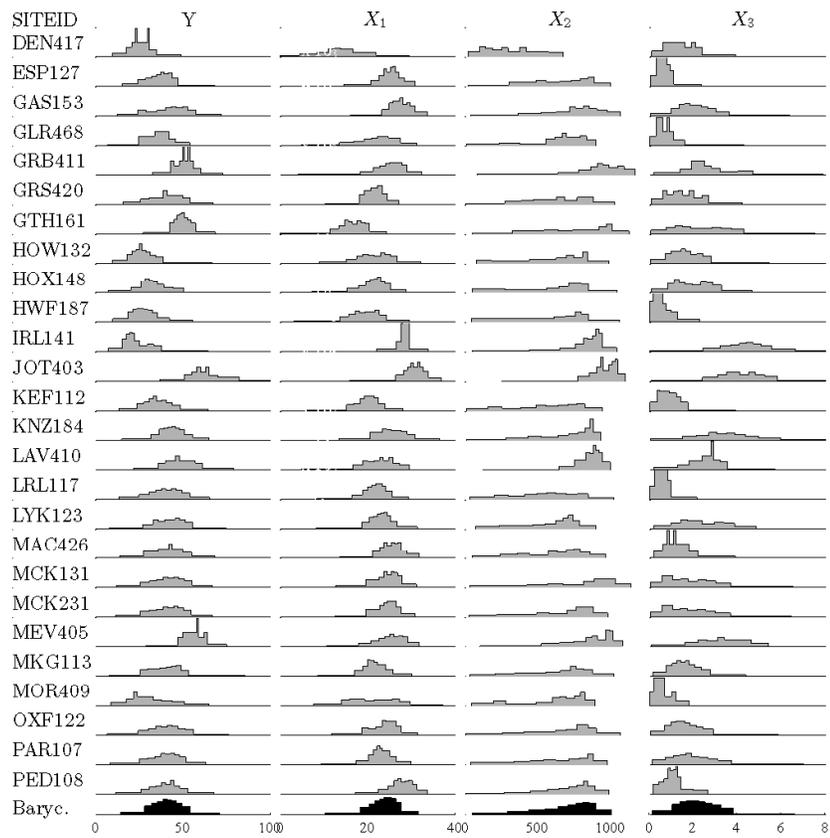}
\caption{Ozone dataset: monitored sites from 27 to 52.}\label{FIG:OZONE2}
\end{figure}

\begin{figure}
\includegraphics[width=0.9\textwidth]{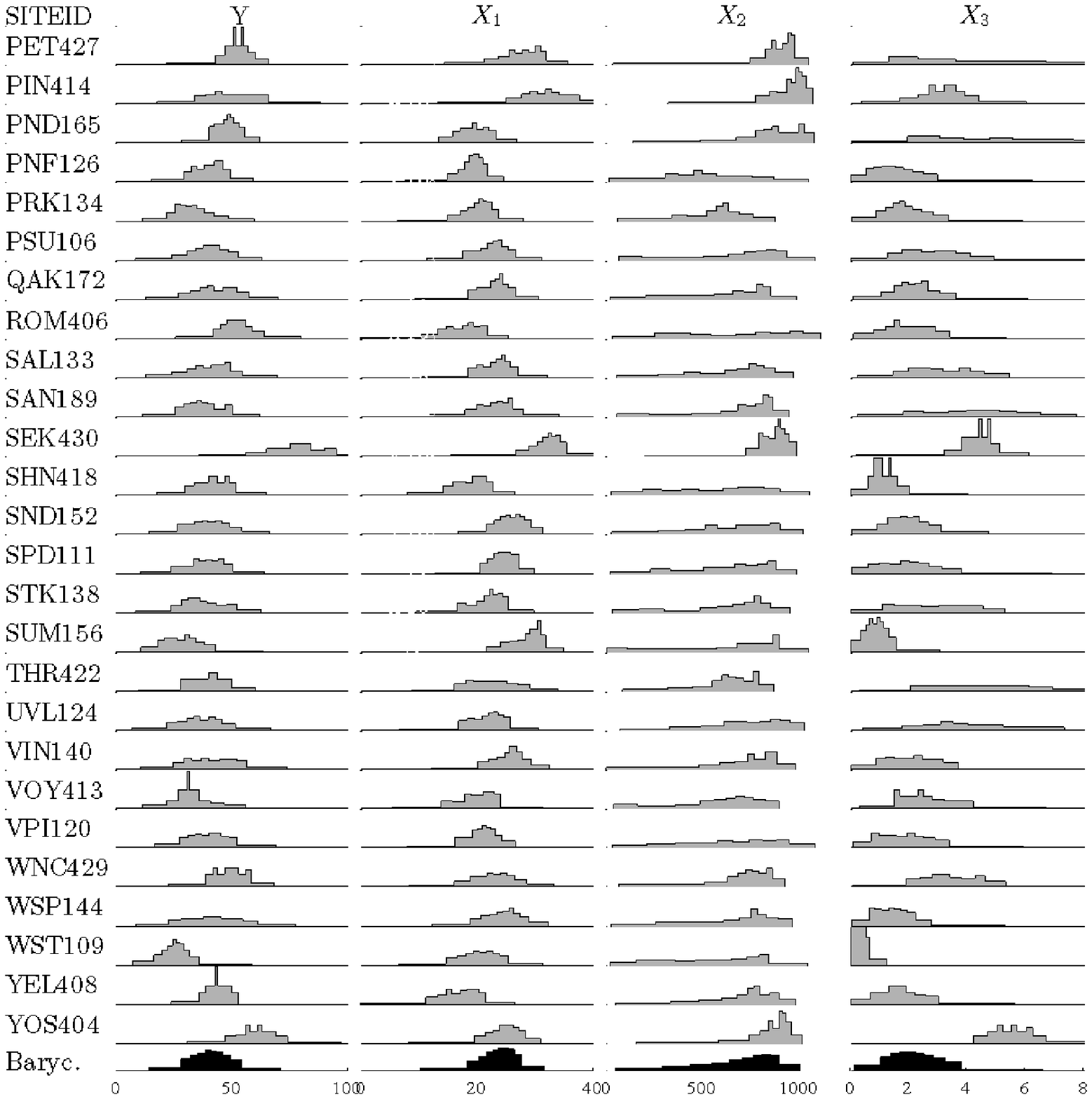}
\caption{Ozone dataset: monitored sites from 53 to 78.}\label{FIG:OZONE3}
\end{figure}

\begin{figure}
\centering{}\includegraphics[scale=0.7]{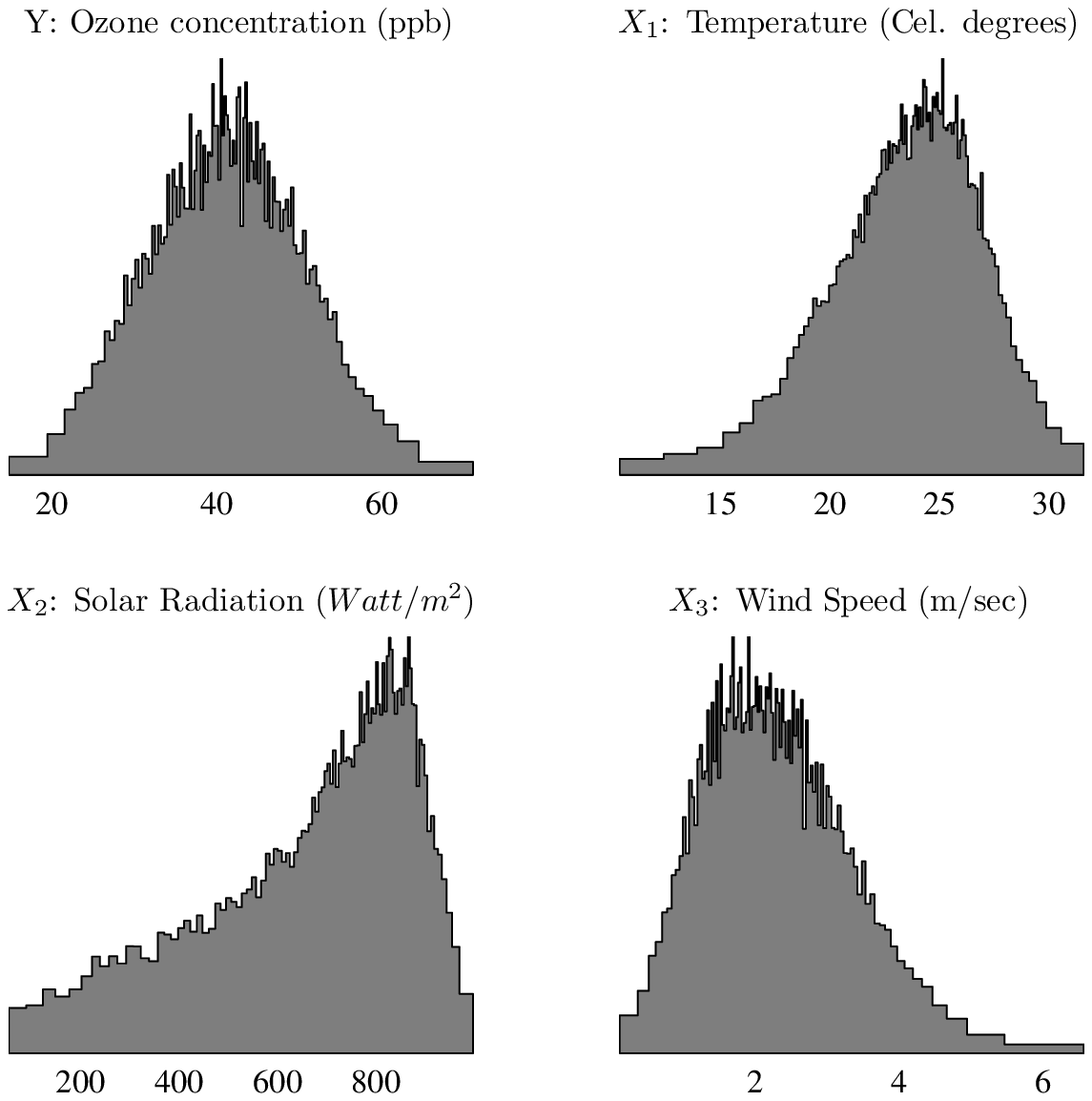}
\caption{Ozone dataset. Barycenters}
\label{Fig: OZO_barycenters}
\end{figure}

Given the distribution of Temperature ($X_{1}$) (Celsius degrees),
the distribution of Solar Radiation ($X_{2}$) (Watts per square meter)
and the distribution of Wind Speed ($X_{3}$) (meters per second),
the main objective is to predict the distribution of Ozone Concentration
($Y$) (Particles per billion) using a linear model. CASTNET collect
hourly data and as period of observation we choose the summer season
of 2010 and the central hours of the days (10 a.m. - 5
p.m.).\\
For each sites we have collected the NPSD of the four variables in
terms of histograms and in figs. \ref{FIG:OZONE1}, \ref{FIG:OZONE2}
and \ref{FIG:OZONE3}
we present the dataset in a graphical way \footnote{We supply the full table of histogram data, the Matlab routines and the workspaces as supplementary material.}. In order to have a visual reference, at the bottom of each figure are shown the barycenters (the mean histogram according to \cite{Irpino20081648}) of each histogram variable. hey appears more in details in  in fig. \ref{Fig: OZO_barycenters}.

In table \ref{TAB:OZO_summarystat} we reported the main summary statistics
for the four histogram variables, while in figure \ref{Fig: OZO_barycenters}
are drawn the four barycenters where it is possible to observe the
average distributions (in the sense of \citet{VER_IR_CP2008})
of the 78 sites for each variable.

\begin{table}
\centering{}{\small }%
\begin{tabular}{lcccc}
\hline
 & {\small Ozone Concentration } & {\small Temperature } & {\small Solar Radiation } & {\small Wind Speed}\\
 & {\small ($Y$ in Ppb)} & {\small ($X_{1}$ in Celsius deg.)} & {\small ($X_{2}$ $Watt/m^{2}$)} & {\small{} ($X_{3}$ m/s)}\\
\hline
{\small Mean (BD)} & {\small 41.2147} & {\small 23.2805} & {\small 645.3507} & {\small 2.3488}\\
{\small Barycenter mean (VI)} & {\small 41.2147} & {\small 23.2805} & {\small 645.3507} & {\small 2.3488}\\
{\small Barycenter std (VI)} & {\small 9.9680} & {\small 3.7641} & {\small 225.7818} & {\small 1.0987}\\
{\small Standard dev. (BD)} & {\small 13.790} & {\small 5.3787} & {\small 252.6736} & {\small 1.7125}\\
{\small Standard dev. (VI)} & {\small 9.5295} & {\small 3.8422} & {\small 113.4308} & {\small 1.1337}\\
\hline
\end{tabular}{\small }\\
{\small }%
\begin{tabular}{cccccccc}
\multicolumn{8}{c}{{\small Correlations}}\\
\hline
 & \multicolumn{3}{c}{{\small Billard-Diday}} &  & \multicolumn{3}{c}{{\small Verde-Irpino}}\\
 & {\small $X_{1}$} & {\small $X_{2}$} & {\small $X_{3}$} &  & {\small $X_{1}$} & {\small $X_{2}$} & {\small $X_{3}$}\\
\hline
{\small $Y$} & {\small 0.2328} & {\small 0.4064} & {\small 0.2951} &  & {\small 0.2473} & {\small 0.6392} & {\small 0.4020}\\
{\small $X_{1}$} &  & {\small 0.2622} & {\small 0.0621} &  &  & {\small 0.4537} & {\small 0.1429}\\
{\small $X_{2}$} &  &  & {\small 0.3013} &  &  &  & {\small 0.4394}\\
\hline
\end{tabular}\caption{Ozone dataset: summary statistics}\label{TAB:OZO_summarystat}
\end{table}

We can note, for example, the different skewness of the barycenters,
in general when the barycenter is skew, we may observe that the NPSD
are in general skew in the same direction. This is not in general
true for symmetric barycenters, that can be generated both from left
and right skewed distributions.

Using the full dataset we estimated the three models and the associated
goodness of fit diagnostics as reported in tables \ref{TAB:Ozone Bill Did}, \ref{TAB: Ozone Dias Brito} and \ref{TAB: Ozone Irpino Verde}.

\begin{table}
\begin{centering}
\begin{tabular}{cccccccr}
\multicolumn{8}{c}{Billard-Diday Model:}\\
\multicolumn{8}{c}{ $\hat{y}_{i}=\hat{\beta_{0}}+\hat{\beta}_{1}x_{i1}+\hat{\beta}_{2}x_{i2}+\hat{\beta}_{3}x_{i3}$}\\
\cline{2-8}
 & \multicolumn{4}{c}{Model parameters} & \multicolumn{3}{c}{Goodness of fit }\\
\cline{2-8}
 & %
$\hat{\beta_{0}}$
 & %
$\hat{\beta}_{1}$
 & %
$\hat{\beta}_{2}$
 & %
$\hat{\beta}_{3}$
 & $\Omega$ & {\footnotesize $Pseudo-R^{2}$} & {\footnotesize $RMSE_{W}$}\\
\cline{2-8}
{\small Observed} & {\small 18.28} & {\small 0.357} & {\small 0.017} & {\small 1.550} & \emph{\footnotesize 0.203} & \emph{\footnotesize 0.024} & \emph{\footnotesize 9.413}\\
 & \multicolumn{7}{c}{{\small Bootstrap estimates}}\\
{\small Mean} & {\small 18.96} & {\small 0.323} & {\small 0.018} & {\small 1.463} & \emph{\footnotesize 0.215} & \emph{\footnotesize 0.103} & \emph{\footnotesize 9.274}\\
{\small Bias} & {\small 0.678} & {\small -0.034} & {\small 0.001} & {\small{} -0.087} &  &  & \\
{\small SE} & {\small 5.077} & {\small 0.182} & {\small 0.005 } & {\small 0.652} &  &  & \\
{\small 2.5\%} & {\small 9.52} & {\small -0.014} & {\small 0.007} & {\small 0.147} & \emph{\footnotesize 0.070} & \emph{\footnotesize 0.000} & \emph{\footnotesize 7.766}\\
{\small 97.5\%} & {\small 28.89} & {\small 0.697} & {\small 0.027} & {\small 2.836} & \emph{\footnotesize 0.350} & \emph{\footnotesize 0.372} & \emph{\footnotesize 10.967}\\
\cline{2-8}
\end{tabular}
\end{centering}
\caption{Ozone dataset: BIllard-Diday model parameters estimated on the full
dataset and bootstrapping the dataset.}
\label{TAB:Ozone Bill Did}
\end{table}

\begin{table}
\begin{centering}
\begin{tabular}{ccccccccccc}
\multicolumn{11}{c}{Dias-Brito model:}\\
\multicolumn{11}{c}{ $\hat{y}_{i}(t)=\hat{\beta_{0}}+\hat{\beta}_{1}x_{i1}(t)+\hat{\beta}_{2}x_{i2}(t)+\hat{\beta}_{3}x_{i3}(t)+\hat{\tilde{\beta}}_{1}\tilde{x}_{i1}(t)+\hat{\tilde{\beta}}_{2}\tilde{x}_{i2}(t)+\hat{\tilde{\beta}}_{3}\tilde{x}_{i3}(t)$}\\
\cline{2-11}
 & \multicolumn{7}{c}{Model parameters} & \multicolumn{3}{c}{Goodness of fit }\\
\cline{2-11}
 & $\hat{\beta_{0}}$ & $\hat{\beta}_{1}$ & $\hat{\beta}_{2}$ & $\hat{\beta}_{3}$ & $\hat{\tilde{\beta}}_{1}$ & $\hat{\tilde{\beta}}_{2}$ & $\hat{\tilde{\beta}}_{3}$ & $\Omega$ & {\footnotesize $Pseudo-R^{2}$} & {\footnotesize $RMSE_{W}$}\\
\cline{2-11}
{\small Observed} & {\small 13.32} & {\small 0.000} & {\small 0.037} & {\small 1.691} & {\small 0.000} & {\small 0.000} & {\small 0.000} & \emph{\footnotesize 0.670} & \emph{\footnotesize 0.371} & \emph{\footnotesize 7.557}\\
 & \multicolumn{10}{c}{{\small Bootstrap estimates}}\\
{\small Mean} & {\small 14.22} & {\small 0.117 } & {\small 0.034 } & {\small 1.709 } & {\small 0.080} & {\small 0.000} & {\small 0.002} & \emph{\footnotesize 0.712} & \emph{\footnotesize 0.358} & \emph{\footnotesize 7.368}\\
{\small Bias} & {\small 0.905} & {\small 0.117} & {\small -0.003} & {\small 0.018} & {\small 0.080} & {\small 0.000} & {\small{} 0.002} &  &  & \\
{\small SE} & {\small 4.760} & {\small 0.161} & {\small 0.004} & {\small 0.610} & {\small 0.112} & {\small 0.000} & {\small{} 0.025} &  &  & \\
{\small 2.5\%} & {\small 5.409} & {\small 0.000} & {\small 0.026} & {\small 0.602} & {\small 0.000} & {\small 0.000} & {\small 0.000} & \emph{\footnotesize 0.625} & \emph{\footnotesize 0.220} & \emph{\footnotesize 5.614}\\
{\small 97.5\%} & {\small 24.46} & {\small 0.540} & {\small 0.040} & {\small 3.070} & {\small 0.391} & {\small 0.000} & {\small 0.000} & \emph{\footnotesize 0.801} & \emph{\footnotesize 0.498} & \emph{\footnotesize 9.126}\\
\cline{2-11}
\end{tabular}
\end{centering}
\caption{Ozone dataset: Dias-Brito model parameters estimated on the full dataset
and bootstrapping the dataset.}
\label{TAB: Ozone Dias Brito}
\end{table}

\begin{table}
\begin{centering}
\begin{tabular}{ccccccccccc}
\multicolumn{11}{c}{Irpino-Verde model:}\\
\multicolumn{11}{c}{ $\hat{y}_{i}(t)=\hat{\beta_{0}}+\hat{\beta}_{1}\bar{x}_{i1}+\hat{\beta}_{2}\bar{x}_{i2}+\hat{\beta}_{3}\bar{x}_{i3}+\hat{\gamma}_{1}x_{i1}^{c}(t)+\hat{\gamma}_{2}x_{i2}^{c}(t)+\hat{\gamma}_{3}x_{i3}^{c}(t)$}\\
\cline{2-11}
 & \multicolumn{7}{c}{Model parameters} & \multicolumn{3}{c}{Goodness of fit }\\
\cline{2-11}
 & %
$\hat{\beta_{0}}$
 & %
$\hat{\beta}_{1}$
 & %
$\hat{\beta}_{2}$
 & %
$\hat{\beta}_{3}$
 & %
$\hat{\gamma}_{1}$
 & %
$\hat{\gamma}_{2}$
 & %
$\hat{\gamma}_{3}$
 & $\Omega$ & {\footnotesize $Pseudo-R^{2}$} & {\footnotesize $RMSE_{W}$}\\
\cline{2-11}
{\small Observed} & 2.928 & -0.346 & 0.070 & 0.395 & 0.915 & 0.018 & 1.887 & \emph{\footnotesize 0.742} & \emph{\footnotesize 0.460} & \emph{\footnotesize 6.999}\\
 & \multicolumn{10}{c}{{\small Bootstrap estimates}}\\
{\small Mean} & 3.108 & -0.353 & 0.070 & 0.363 & 0.928 & 0.018 & 1.958 & \emph{\footnotesize 0.758} & \emph{\footnotesize 0.474} & \emph{\footnotesize 6.729}\\
{\small Bias} & 0.181 & -0.008 & 0.000 & -0.032  & 0.013  & 0.000 &  0.071 &  &  & \\
{\small SE} & 7.180 & 0.271 & 0.010 & 0.823 & 0.237 & 0.003  & 0.542 &  &  & \\
{\small 2.5\%} & -11.24 & -0.846 & 0.052 & -1.186 & 0.482 & 0.012 & 1.054 & \emph{\footnotesize 0.675} & \emph{\footnotesize 0.296} & \emph{\footnotesize 5.267}\\
{\small 97.5\%} & 18.87 & 0.173 & 0.090 & 2.030 & 1.377 & 0.024 & 3.115 & \emph{\footnotesize 0.829} & \emph{\footnotesize 0.625} & \emph{\footnotesize 8.298}\\
\cline{2-11}
\end{tabular}
\par\end{centering}
\caption{Ozone dataset: Irpino-Verde model parameters estimated on the full
dataset and bootstrapping the dataset.}
\label{TAB: Ozone Irpino Verde}
\end{table}

Also in this case, we performed a bootstrap estimates of the parameters
and of the goodness of fit measures of the three models in tables
\ref{TAB:Ozone Bill Did}, \ref{TAB: Ozone Dias Brito} and \ref{TAB: Ozone Irpino Verde}.

Observing the goodness of fit measures of the three models we can
conclude that the Irpino-Verde and the Dias-Brito model fit better
the linear regression relationship than the Billard-Diday model, and
the Irpino-Verde model is slightly more accurate than the Dias-Brito
one. Also in this case, the Irpino-Verde model parameters give an easier interpretation. Reading
the Irpino-Verde bootstrapped model, we may assert that the Ozone
concentration distribution of a site depends from the mean \emph{solar
radiation} where for each $\Delta Watt/m^{2}$ a $0.070(ppb)$
variation of the \emph{Ozone concentration} mean level it is expected, while in general
we cannot say that the mean level of \emph{temperature} and of \emph{wind
speed} induces a significant variation of the \emph{ozone concentration
level }(the $95\%$ bootstrap confidence intervals include zero).
Furthermore, we may say that the variability of the \emph{ozone concentration}
is quite the same as the \emph{temperature }($0.928$), a unit variation
in the variability of the \emph{solar radiation} induces a variation
of $0.018(ppb)$ and a variation of the variability of the \emph{Wind
Speed} causes an increase in the variability of $1.958(ppb)$. Similar
conclusions can be derived reading the Dias-Brito model even if it gives a different interpretation of the parameter associated with the symmetric histogram variables.

\section{Conclusions}

The paper present a novel linear regression technique for  data described by probability-like distributions, using their quantile functions and the ordinary least squares method based on the Wasserstein distance. Considering
the nature of the data we proposed to use a particular decomposition
of the Wasserstein distance for the definition of the regression model.
We have also furnished an alternative goodness of fit index which takes into account the differences in shape and size of the quantile distributions of the independent variables.
The proposed model allow a better interpretation of the parameters and, in the multivariate case, showed better fit to the data with respect to the two main approaches
presented in the literature. Further it allows an easier interpretation
of the results. We also showed that the method can be used with a
variety of numerical probabilistic symbolic data. Considering the
complexity of the error term, the
classic parameter inferential properties can not straightforward be extended to the regression of NPSD.
We consider to address new efforts in the direction of investigate
the properties of the involved estimators.

\appendix

\section{OLS solution details\label{sec:OLS-solution-details}}

In this appendix we show main passages for the solution of the LS problem that leads to Eq. (\ref{eq:slopes}) and (\ref{eq:GAMMA}). 
\begin{equation}
\begin{array}{l}
SSE\mathbf{(B,\Gamma)=e^{T}e}=\\
=\left[{\bf Y}-{\bf \bar{X}}_{+}{\bf B}-{\bf X}^{{\bf c}}{\bf \Gamma}\right]^{T}\left[{\bf Y}-{\bf \bar{X}}_{+}{\bf B}-{\bf X}^{{\bf c}}{\bf \Gamma}\right]=\\
={\bf Y}^{T}{\bf Y}-{\bf Y}^{T}{\bf \bar{X}}_{+}{\bf B}-{\bf Y}^{T}{\bf X}^{{\bf c}}{\bf \Gamma}-{\bf B}^{T}{\bf \bar{X}}{}_{+}^{T}{\bf Y}+{\bf B}^{T}{\bf \bar{X}}_{+}^{T}+{\bf \bar{X}B}+\\
+{\bf B}^{T}{\bf \bar{X}}_{+}^{T}{\bf X}^{{\bf c}}{\bf \Gamma}-{\bf \Gamma}^{T}{\bf X}^{{\bf c}}{}^{T}{\bf Y}+{\bf \Gamma}^{T}{\bf X}^{{\bf c}}{}^{T}{\bf \bar{X}B}+{\bf \Gamma}^{T}{\bf X}^{{\bf c}}{}^{T}{\bf X}^{{\bf c}}{\bf \Gamma}=\\
={\bf Y}^{T}{\bf Y}-2{\bf B}^{T}{\bf \bar{X}}_{+}^{T}{\bf Y}+{\bf B}^{T}{\bf \bar{X}}_{+}^{T}{\bf \bar{X}B}+2{\bf B}^{T}{\bf \bar{X}}_{+}^{T}{\bf X}^{{\bf c}}{\bf \Gamma}+\\
-2{\bf \Gamma}^{T}{\bf X}^{{\bf c}}{}^{T}{\bf Y}+{\bf \Gamma}^{T}{\bf X}^{{\bf c}}{}^{T}{\bf X}^{{\bf c}}{\bf \Gamma}
\end{array}
\end{equation}
 first order conditions
\[
\begin{array}{l}
\frac{\delta SSE}{\delta B}=-2{\bf \bar{X}}_{+}^{T}{\bf Y}+2{\bf \bar{X}}_{+}^{T}{\bf \bar{X}}_{+}{\bf B}+2{\bf \bar{X}}_{+}^{T}{\bf X}^{{\bf c}}{\bf \Gamma}=0\\
\frac{\delta SSE}{\delta\Gamma}=+2{\bf B}^{T}{\bf \bar{X}}_{+}^{T}{\bf X}^{{\bf c}}-2{\bf X}^{{\bf cT}}{\bf Y}+2{\bf X}^{{\bf cT}}{\bf X}^{{\bf c}}{\bf \Gamma}=0
\end{array}
\]

\[
\frac{\delta SS}{\delta{\bf B}}=0\to-2\underbrace{{\bf \bar{X}}_{+}^{T}{\bf Y}}_{{\bf \bar{X}}_{+}^{T}{\bf \bar{Y}}}+2{\bf \bar{X}}_{+}^{T}{\bf \bar{X}}_{+}{\bf B}+2\underbrace{{\bf \bar{X}}_{+}^{T}{\bf X}^{{\bf c}}}_{0}{\bf \Gamma}=0\to{\bf B}=\left({\bf \bar{X}}_{+}^{T}{\bf \bar{X}}_{+}\right)^{-1}{\bf \bar{X}}_{+}^{T}{\bf \bar{Y}}
\]

\[
\frac{\delta SS}{\delta\Gamma}=0\to2{\bf B}^{T}\underbrace{{\bf \bar{X}}_{+}^{T}{\bf X}^{{\bf c}}}_{0}-2\underbrace{{\bf X}^{{\bf cT}}{\bf Y}}_{{\bf X}^{{\bf c}}{}^{T}{\bf Y}^{{\bf c}}}+2{\bf X}^{{\bf c}}{}^{T}{\bf X}^{{\bf c}}{\bf \Gamma}=0\to{\bf \Gamma}=\left({\bf X}^{{\bf c}}{}^{T}{\bf X}^{{\bf c}}\right)^{-1}{\bf X}^{{\bf c}}{}^{T}{\bf Y}^{{\bf c}}
\]

\section{The decomposition of the sum of square of Y \label{sec:Sum-of-square}}

Considering $\mathbf{1}=[1]_{n\times1}$, $SSY$ can be written as:

\[
\begin{array}{l}
SSY=n\cdot s_{y}^{2}=\sum\limits _{i=1}^{n}d_{W}^{2}\left(y_{i}(t),\bar{y}(t)\right)=\sum\limits _{i=1}^{n}\int\limits _{0}^{1}\left[y_{i}(t)-\bar{y}(t)\right]^{2}dt=\\
=\left(\underbrace{{\bf \bar{Y}}+{\bf Y}^{c}}_{{\bf Y}}-{\bf 1}\underbrace{\left(\bar{y}+\bar{y}^{c}(t)\right)}_{\bar{y}(t)}\right)^{T}\left(\underbrace{{\bf \bar{Y}}+{\bf Y}^{c}}_{{\bf Y}}-{\bf 1}\underbrace{\left(\bar{y}+\bar{y}^{c}(t)\right)}_{\bar{y}(t)}\right)
\end{array}
\]

further

\[
\begin{array}{l}
SSY=\int\limits _{0}^{1}\left({\bf Y}-{\bf 1}\bar{y}(t)+{\bf \hat{Y}}-{\bf \hat{Y}}\right)^{T}\left({\bf Y}-{\bf 1}\bar{y}(t)+{\bf \hat{Y}}-{\bf \hat{Y}}\right)dt=\\
=\int\limits _{0}^{1}\left(\underbrace{{\bf Y}-{\bf \hat{Y}}}_{{\bf {\varepsilon}}(t)}-\left({\bf 1}\bar{y}(t)-{\bf \hat{Y}}(t)\right)\right)^{T}\left(\underbrace{{\bf Y}(t)-{\bf \hat{Y}}(t)}_{{\bf {\varepsilon}}(t)}-\left({\bf 1}\bar{y}(t)-{\bf \hat{Y}}(t)\right)\right)dt=\\
=\int\limits _{0}^{1}\left(\underbrace{{\bf \varepsilon}(t)^{T}{\bf \varepsilon}(t)}_{SSE}+\underbrace{\left({\bf 1}\bar{y}(t)-{\bf \hat{Y}}(t)\right)^{T}\left({\bf 1}\bar{y}(t)-{\bf \hat{Y}}(t)\right)}_{SSR}-2\left({\bf 1}\bar{y}(t)-{\bf \hat{Y}}(t)\right)^{T}{\bf \varepsilon}(t)\right)dt=\\
=SSE+SSR-2\int\limits _{0}^{1}\left({\bf 1}\bar{y}(t)-{\bf \hat{Y}}(t)\right)^{T}{\bf \varepsilon}(t)dt=\\
=SSE+SSR-2\left[\underbrace{\int\limits _{0}^{1}\left({\bf 1}\bar{y}(t)\right)^{T}{\bf \varepsilon}(t)dt}_{(I)}-\underbrace{\int\limits _{0}^{1}\left({\bf \hat{Y}}(t)\right)^{T}{\bf \varepsilon}(t)dt}_{(II)}\right]=
\end{array}
\]

for (I) we have

\[
\begin{array}{l}
\underbrace{\int\limits _{0}^{1}\left({\bf 1}_{n\times1}\bar{y}(t)\right)^{T}{\bf \varepsilon}(t)dt}_{(I)}=\int\limits _{0}^{1}\left({\bf 1}\bar{y}(t)\right)^{T}\left({\bf Y}(t)-{\bf \bar{X}_{+}B}-{\bf X}^{c}(t){\bf \Gamma}\right)dt=\\
=\int\limits _{0}^{1}\left(\bar{y}(t)\right)^{T}{\bf 1}^{T}{\bf Y}(t)dt-\int\limits _{0}^{1}\left(\bar{y}(t)\right)^{T}{\bf 1}^{T}{\bf \bar{X}_{+}B}dt-\int\limits _{0}^{1}\left(\bar{y}(t)\right)^{T}{\bf 1}^{T}{\bf X}^{c}(t){\bf \Gamma}dt=\\
=\int\limits _{0}^{1}\left(\bar{y}(t)\right)^{T}n\cdot\bar{y}(t)dt-n\bar{y}^{2}-\int\limits _{0}^{1}\left(\bar{y}(t)\right)^{T}n{\bf \bar{X}}^{c}(t){\bf \Gamma}dt=n\cdot\sigma_{\bar{y}}^{2}+n\bar{y}^{2}-n\bar{y}^{2}-n\sum\limits _{i=1}^{p}\gamma_{p}r_{\bar{y}\bar{x}_{j}}\sigma_{\bar{y}}\sigma_{\bar{x}_{j}}=\\
=n\cdot\sigma_{\bar{y}}^{2}-n\sum\limits _{i=1}^{p}\gamma_{p}r_{\bar{y}\bar{x}_{j}}\sigma_{\bar{y}}\sigma_{\bar{x}_{j}}=n\cdot\left(\sigma_{\bar{y}}^{2}-\sum\limits _{i=1}^{p}\gamma_{p}r_{\bar{y}\bar{x}_{j}}\sigma_{\bar{y}}\sigma_{\bar{x}_{j}}\right)
\end{array}
\]

for (II) we have

\[
\begin{array}{l}
-\underbrace{\int\limits _{0}^{1}\left({\bf \hat{Y}}(t)\right)^{T}{\bf \varepsilon}(t)dt}_{(II)}=-\int\limits _{0}^{1}\left({\bf \bar{X}_{+}B}+{\bf X}^{c}(t){\bf \Gamma}\right)^{T}\left({\bf Y}(t)-{\bf \bar{X}_{+}B}-{\bf X}^{c}(t){\bf \Gamma}\right)dt=\\
=-\int\limits _{0}^{1}\left({\bf \bar{X}_{+}B}\right)^{T}{\bf Y}(t)dt+\int\limits _{0}^{1}\left({\bf \bar{X}_{+}B}\right)^{T}{\bf \bar{X}_{+}B}dt+\int\limits _{0}^{1}\left({\bf \bar{X}_{+}B}\right)^{T}{\bf X}^{c}(t){\bf \Gamma}dt+\\
-\int\limits _{0}^{1}\left({\bf X}^{c}(t){\bf \Gamma}\right)^{T}\left({\bf Y}(t)\right)dt+\int\limits _{0}^{1}\left({\bf X}^{c}(t){\bf \Gamma}\right)^{T}\left({\bf \bar{X}B}\right)dt+\int\limits _{0}^{1}\left({\bf X}^{c}(t){\bf \Gamma}\right)^{T}\left({\bf X}^{c}(t){\bf \Gamma}\right)dt=\\
=\underbrace{-{\bf \bar{Y}}\left({\bf \bar{X}B}\right)^{T}+\left({\bf \bar{X}B}\right)^{T}\left({\bf \bar{X}B}\right)}_{{\bf {B}}\nabla SSE({\bf {B}})=0}+\underbrace{\left({\bf \bar{X}B}\right)^{T}\int\limits _{0}^{1}\left({\bf X}^{c}(t){\bf \Gamma}\right)dt}_{0}+\\
-\int\limits _{0}^{1}\left({\bf X}^{c}(t){\bf \Gamma}\right)^{T}\left({\bf Y}(t)\right)dt+\underbrace{\int\limits _{0}^{1}\left({\bf X}^{c}(t){\bf \Gamma}\right)^{T}\left({\bf \bar{X}B}\right)dt}_{0}+\int\limits _{0}^{1}\left({\bf X}^{c}(t){\bf \Gamma}\right)^{T}\left({\bf X}^{c}(t){\bf \Gamma}\right)dt=\\
=-\int\limits _{0}^{1}\left({\bf X}^{c}(t){\bf \Gamma}\right)^{T}\left({\bf Y}(t)\right)dt+\int\limits _{0}^{1}\left({\bf X}^{c}(t){\bf \Gamma}\right)^{T}\left({\bf X}^{c}(t){\bf \Gamma}\right)dt=\\
=-\int\limits _{0}^{1}\left({\bf X}^{c}(t){\bf \Gamma}\right)^{T}\left({\bf \bar{Y}}+{\bf Y}^{c}(t)\right)dt+\int\limits _{0}^{1}\left({\bf X}^{c}(t){\bf \Gamma}\right)^{T}\left({\bf X}^{c}(t){\bf \Gamma}\right)dt=\\
=-\underbrace{\int\limits _{0}^{1}\left({\bf X}^{c}(t){\bf \Gamma}\right)^{T}{\bf \bar{Y}}dt}_{0}-\int\limits _{0}^{1}\left({\bf X}^{c}(t){\bf \Gamma}\right)^{T}{\bf Y}^{c}(t)dt+\int\limits _{0}^{1}\left({\bf X}^{c}(t){\bf \Gamma}\right)^{T}\left({\bf X}^{c}(t){\bf \Gamma}\right)dt=\\
=\underbrace{\int\limits _{0}^{1}\left({\bf X}^{c}(t){\bf \Gamma}\right)^{T}\left({\bf X}^{c}(t){\bf \Gamma}\right)dt-\int\limits _{0}^{1}\left({\bf X}^{c}(t){\bf \Gamma}\right)^{T}{\bf Y}^{c}(t)dt}_{=0\quad{\text {if using Ordinary LS solutions}}}={\bf \Gamma}\nabla f({\bf \Gamma})\\
\int\limits _{0}^{1}\left({\bf X}^{c}(t){\bf \Gamma}\right)^{T}\left({\bf X}^{c}(t){\bf \Gamma}\right)dt-\int\limits _{0}^{1}\left({\bf X}^{c}(t){\bf \Gamma}\right)^{T}{\bf Y}^{c}(t)dt=\\
={\bf \Gamma}\underbrace{\int\limits _{0}^{1}\left[\left({\bf X}^{c}(t)\right)^{T}{\bf X}^{c}(t){\bf \Gamma}-\left({\bf X}^{c}(t)\right)^{T}{\bf Y}^{c}(t)\right]dt}_{\nabla SSE({\bf {\Gamma}})}
\end{array}
\]

thus

\[
bias=-2\left[n\cdot\left(\sigma_{\bar{y}}^{2}-\sum\limits _{j=1}^{p}\gamma_{j}r_{\bar{y}\bar{x}_{j}}\sigma_{\bar{y}}\sigma_{\bar{x}_{j}}\right)+{\bf \Gamma}\nabla SSE({\bf \Gamma})\right]
\]

the decomposition is then:
\[
SSY=SSE+SSR-2\left[n\cdot\left(\sigma_{\bar{y}}^{2}-\sum\limits _{j=1}^{p}\gamma_{j}r_{\bar{y}\bar{x}_{j}}\sigma_{\bar{y}}\sigma_{\bar{x}_{j}}\right)+{\bf \Gamma}\nabla SSE({\bf \Gamma})\right]
\]

\bibliographystyle{elsarticle-num-names}
\bibliography{biblioREG}

\end{document}